# Enabling Electrical Readout of Néel vector reversal in a van der Waals Antiferromagnet

Raghvendra Posti[1], Ravi Kumar Bandepalli[1], Wenhao Liu[2], Anshuman Sahoo[1], Pratyush Saud[1], Zixin Zhai[2], Aswin L. N. Kondusamy[2], Zhenhong Cui[1], I-Hsuan Kao[1], Aalok Tiwari[1], Thomas Poirier[3], James H. Edgar[3], Kenji Watanabe[4], Takashi Taniguchi[5], Bing Lv[2,6], Jyoti Katoch[1], and Simranjeet Singh[1,*]

[1]*Department of Physics, Carnegie Mellon University, Pittsburgh, PA, 15213, USA*
[2]*Department of Physics, The University of Texas at Dallas, Richardson, TX 75080, USA*
[3]*Tim Taylor Department of Chemical Engineering, Kansas State University, Manhattan, Kansas 66506, USA*
[4]*Research Center for Electronic and Optical Materials, National Institute for Materials Science, 1-1 Namiki, Tsukuba 305-0044, Japan*
[5]*Research Center for Materials Nanoarchitectonics, National Institute for Materials Science, 1-1 Namiki, Tsukuba 305-0044, Japan*
[6]*Department of Materials Science & Engineering, The University of Texas at Dallas, Richardson, TX 75080 USA*

---

[*]*Email: simranjs@andrew.cmu.edu*

**Owing to its robustness against external perturbations and intrinsically ultrafast dynamics, the Néel vector in antiferromagnets (AFMs) can enable the development of next-generation spintronic and magnonic devices for memory and computing applications. To realize AFM-based magnetic memory devices, one of the key requirements is to demonstrate electrical readout of 180-degree reversal of Néel vector in thin film AFMs, which remains critically missing. In this work, we report experimental demonstration of a novel transport methodology to detect Néel vector reversal in atomically thin films of a van der Waals (vdW) based A-type AFM. For this, we utilize spin-dependent electronic band properties of CrSBr by coupling it to a spin-polarized layer, separated by a tunnel barrier. In this configuration, the spin-dependent tunnelling magnetoresistance (MR) becomes sensitive to the relative orientation between the magnetization of the reference electrode and the interfacial sublattice magnetization of the AFM layer, in turn enabling electrical detection of the Néel vector orientation. Importantly, the observed MR can also reveal 180-degree reversal of Néel vector in even-layers of CrSBr, wherein adjacent sublattice magnetic layers are exactly compensated and the net magnetization vanishes and thus establishes a broadly applicable strategy for electrical detection of Néel vector in vdW-based AFMs.**

Strong exchange interaction in AFMs leads to ultra-fast dynamics of the order parameter, i.e., Néel vector, which is orders of magnitude higher than those observed in ferromagnets (FMs)[1-3]. This can allow for Néel vector to be manipulated on ultrafast timescales, making AFMs highly attractive for faster response spintronic and magnonic devices[2-5]. For instance, to realize AFM-based ultrafast magnetic memory devices, one must demonstrate a pathway to induce energy-efficient Néel vector switching and subsequent detection of the Néel vector by electrical means. Critically, due to absence of net magnetization in AFMs, there are no existing magnetoresistance effects to enable the electrical readout of complete 180-degree reversal of Néel vector in the compensated collinear AFMs. Previously, magnetoresistance effects, including proximity induced phenomena, have enabled electrical readout of in-plane reorientation of the Néel vector by only 90º or 120º in collinear AFMs[6-10]. Recently discovered vdW-based A-type AFMs provide alternative material platform to enable electrical readout and control the Néel vector. A variety of magnetic configurations can be realized in this class of AFM materials, with representative examples including in-plane easy-axis anisotropy in CrSBr[11-20], easy-plane magnetization in $CrCl_3$[21,22], and out-of-plane (perpendicular) easy-axis anisotropy in $CrPS_4$[23,24] and $MnBi_2Te_4$[25], among many others. Moreover, vdW-based magnetic systems display highly intriguing properties, such as thickness-dependent magnetic ground state[26,27], electric field tunability of the ordering temperature[28], enhancement of interlayer AFM exchange coupling in the ultra-thin limit[29,30], and tunable magnon-magnon coupling[22], to name a few. These characteristics provide novel means to control magnetic ordering, which is not possible with traditional magnetic systems.

Importantly, vdW-based semiconducting AFMs, exhibit intrinsic spin-dependent band structures, which can lead to spin-dependent transport properties to enable electrical detection of magnetic order[12,23]. Specifically, the magnetoresistive (MR) effects in CrSBr becomes sensitive to the relative orientation of individual magnetic layers, exhibiting a spin-filtering effect[12,13,16,31]: the resistance is enhanced in the fully AFM state and reduced in a field-induced FM configuration. This spin-dependent transport can be exploited for Néel vector detection by integrating the AFM layer with a tunnel barrier-separated spin-polarized electrode. Here, we experimentally demonstrate electrical readout of 180-degree reversal of Néel vector by employing a tunnel-based device configuration, wherein atomically thin CrSBr layers are integrated with a FM electrode through a hexagonal boron nitride (hBN) tunnel barrier. In this configuration, the spin-dependent vertical tunnelling current becomes sensitive to the relative orientation between the spin polarization of the reference FM electrode and the sublattice magnetization of the CrSBr layer to enable electrical detection 180-degree reversal of Néel vector.

CrSBr is a semiconducting AFM with an orthorhombic crystal structure belonging to the $P_{mmn}$ space group and $D_{2h}$ point group. The magnetic ground state in CrSBr is characterized by A-type AFM order, consisting of ferromagnetically coupled magnetic moments within an individual layer and antiferromagnetic coupling between adjacent layers (crystal schematic shown in Fig. 1a)[11-20]. CrSBr exhibits a strong magneto-crystalline anisotropy, with the magnetic easy axis aligned along the in-plane crystallographic *b*-axis and exhibits weak interlayer exchange coupling. The interlayer exchange field ($H_{ex}$) in CrSBr is typically weaker than the magnetic anisotropy field ($H_{an}$), resulting in a sharp spin-flip (SF) transition that separates the AFM and field-polarized FM states[12,13,16,19,31,32]. The electrical transport in CrSBr, both longitudinal and vertical, is strongly coupled to its magnetic states[11-13,31,33-35]. In the AFM state, where adjacent layers are aligned in antiparallel configuration, the resistance is relatively high. Upon the SF transition into the FM phase, the resistance decreases abruptly due to the reduced bandgap relative to the AFM state[34,36,37]. Consequently, CrSBr exhibits a pronounced spin-filtering transport behaviour (both vertical and in-plane transport), with the antiparallel AFM state corresponding to a high-resistance configuration and the parallel FM state corresponding to a low-resistance configuration, as schematically depicted in Fig. 1a. To establish similar behaviour in our devices, we first measured the longitudinal magnetoresistance of multi-layer CrSBr (~10 nm) encapsulated between h-BN layers (insets to Fig. 1c shows the heterostructure schematic and optical image of the device) as a function of magnetic field applied along the easy axis of CrSBr. The magnetoresistance show in Fig. 1c is defined as:

$$\text{MR (\%)} = \frac{R(B) - R(FM)}{R(FM)} \times 100$$

where R(B) denote the device resistance measured at magnetic field B, and R(FM) is the average resistance in the fully field-polarized FM state at maximum positive and negative magnetic fields. In the AFM phase, multilayer CrSBr consists of two oppositely aligned magnetic sublattices, $\boldsymbol{m_A}$ (red arrow) and $\boldsymbol{m_B}$ (blue arrow), with order parameter is described by the Néel vector, $\widehat{\boldsymbol{N}} = \frac{\boldsymbol{m_A} - \boldsymbol{m_B}}{2}$ (pink arrow). As the magnetic field increases, the SF transition of the sublattices gives rise to an abrupt resistance change near ∼350 mT, separating the high-resistance AFM state around low fields from the field-induced low-resistance FM state. Similarly, vertical transport measurements in a bilayer CrSBr device in Fig 1d (side-view schematic of the device and measurement scheme are shown in the inset of Fig. 1d) exhibit a sharp SF transition near ~200 mT, separating the high-resistance AFM state from the low-resistance FM state. The reduced SF field in the bilayer device is attributed to the weaker interlayer AFM exchange interaction. This behaviour persists up to ~130 K, consistent with reported Néel temperature ($T_N$) of CrSBr.

Importantly, we note that despite its sensitivity to the field-driven AFM-FM transition, the spin-filter MR remains insensitive to a 180-degree reversal of the magnetic order, as illustrated schematically in Fig. 1a. In other words, previously reported MR measurements in CrSBr cannot distinguish between Néel vector states of opposite polarity. Nevertheless, both longitudinal and vertical transport measurements establish that the resistance strongly depends on the relative magnetic alignment between adjacent CrSBr layers, indicating that charge transport in CrSBr is highly spin dependent. The spin-filter tunnelling effect in CrSBr can be understood as spin-dependent interlayer hopping of charge carriers between adjacent ferromagnetic monolayers across the van der Waals gap[26,33,38,39]. This behaviour originates from the spin-polarized electronic band structure of individual CrSBr layers, leading to spin-selective tunnelling between neighbouring layers.

To enable electrical detection 180-degree reversal of Néel vector, we propose a detection scheme as illustrated in Fig. 1b. Conventional transition metal FMs, such as cobalt (Co), possess exchange-split majority- and minority-spin bands[40,41]. A tunnel junction formed between a transition metal FM and CrSBr, separated by a thin tunnel barrier, can potentially enable electrical detection of the Néel vector through spin-dependent tunnelling between the FM and the adjacent CrSBr layer. In Fig. 1b, CrSBr in ground state, consisting of two oppositely aligned magnetic sublattices, $\boldsymbol{m_A}$ (red) and $\boldsymbol{m_B}$ (blue) forming the Néel vector, $\widehat{\boldsymbol{N}}$ (purple arrow), is separated from a conventional transition-metal FM (magnetization shown by the green arrow) by a thin tunnel barrier. In such a tunnel junction, the resistance depends on the relative alignment between the FM magnetization and the interfacial CrSBr sublattice ($\boldsymbol{m_A}$, here), and consequently on the orientation of the Néel vector. Depending on whether FM magnetization and the Néel vector are aligned parallel (or antiparallel) to each other, the MR response can yield a low (or high) resistance state. We note that the tunnelling MR in vertical devices can have a pronounced bias dependence, as it is observed in our devices and is discussed later[42-45]. Depending on the applied bias, either the low- or high-resistance state may correspond to parallel or antiparallel alignment between the FM magnetization and the Néel

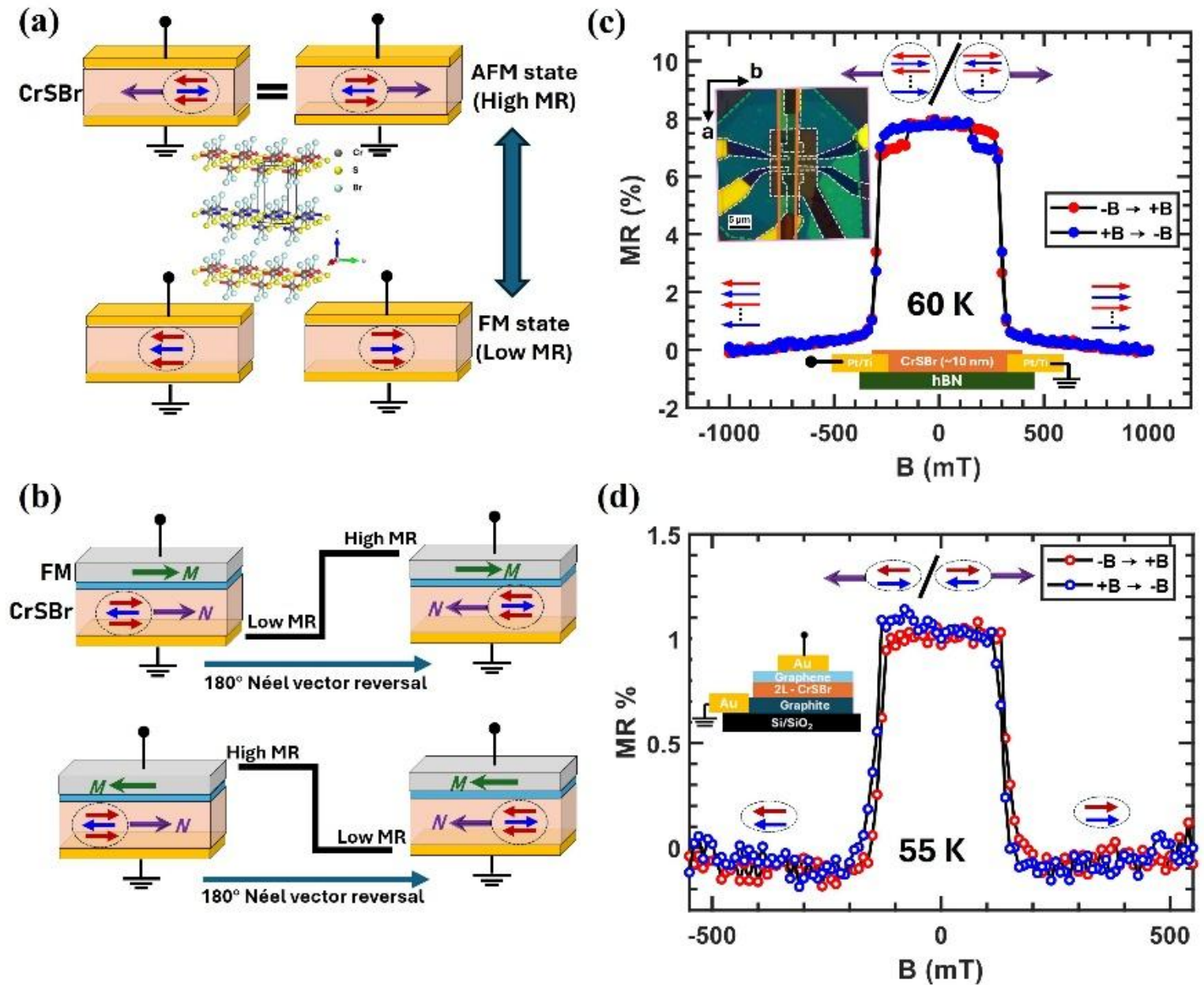

**Fig. 1. Spin-dependent transport in CrSBr. (a)** Schematic of the A-type AFM order in CrSBr (center). Each monolayer exhibits intralayer ferromagnetic coupling, while adjacent layers are coupled antiferromagnetically. In the AFM state, the two sublattices with magnetization vectors $m_A$ and $m_B$ (shown by red and blue arrows, respectively), are antiferromagnetically aligned, giving rise to a Néel vector, $\widehat{\boldsymbol{N}} = \frac{\boldsymbol{m_A} - \boldsymbol{m_B}}{2}$ (purple arrow). The two degenerate AFM configurations with 180°-reversed Néel vectors correspond to high-resistance states. In contrast, the field-induced transition to either of the two possible ferromagnetic alignments of $\boldsymbol{m_A}$ and $\boldsymbol{m_B}$ corresponds to low-resistance states, demonstrating the spin-filtering nature of the system. **(b)** Relative orientation of the magnetization (green arrow) of a FM layer and the CrSBr sublattice magnetizations (red and blue arrows), defining the Néel vector (purple arrow), and the corresponding low- and high-resistance states for all four possible configurations. **(c)** Longitudinal MR as a function of magnetic field applied along b-axis of CrSBr, highlighting the spin-filtering effect. The FM state above the spin-flip transition exhibits lower MR than the AFM state below the transition. Insets: Side-view illustration of the heterostructure and measurement geometry, alongside an optical micrograph of the fabricated device. The outlines of the individual flakes follow the same color convention used in the schematic. **(d)** Vertical MR of bilayer CrSBr measured as a function of magnetic field along b-axis of CrSBr, revealing spin-dependent transport in CrSBr. Insets: Side-view schematic of the heterostructure and measurement geometry

vector. Nevertheless, irrespective of this bias-dependent inversion, such vertical heterostructures should exhibit a robust two-level MR response, enabling electrical readout of the Néel vector orientation with respect to the FM reference layer.

To experimentally demonstrate this mechanism to detect 180-degree reversal of Néel vector, we fabricate vertical tunnel junction device based on Co/hBN/CrSBr/graphite heterostructure. The side-view schematic of such a device is shown in Fig. 2a. Atomically thin flakes of graphite, trilayer (3L) CrSBr, and bilayer (2L) hBN were mechanically exfoliated and sequentially assembled on a thermally oxidized Si/$SiO_2$ substrate using a dry-transfer technique (see Methods for details). The thickness of each constituent flake was identified from optical contrast and independently confirmed by atomic force microscopy measurements. The top panel of Fig. 2b shows an optical image of the 3L-CrSBr flake, with the active device region outlined. The Co electrode and other electrical contacts were subsequently defined using standard nanofabrication procedures. In this device geometry, Co acts as the FM electrode owing to its spin-polarized electronic density of states. The bottom panel of Fig. 2b shows an optical image of the final fabricated device (Device A) with multiple top contacts (A1, A2, and A3) to perform independent measurements. The top contacts A1, A2, and A3 are 400 nm-wide Co electrode, a 450 nm-wide Au electrode, and a 450 nm-wide Co electrode, respectively. The 400 nm-wide Co contact (A1) is hereafter referred to as Device A1. The I-V characteristics of vertical transport through Device A1 exhibits a strongly nonlinear behaviour (Supplementary Fig. S1) that may reflect contributions from both the semiconducting CrSBr layer and the hBN tunnel barrier. Notably, vertical transport through atomically thin CrSBr layers in the absence of a tunnel barrier is typically characterized by a nearly linear I-V response due to direct tunnelling between electrodes[13]. Therefore, the observed nonlinear transport behaviour indicates the formation of a high-quality tunnel junction in our heterostructure device. Next, we measure vertical MR response by applying magnetic field along the *b*-axis of CrSBr (Fig. 2b). The measured MR response, as shown in Fig. 2c, follows characteristic behaviour of CrSBr, showing saturation in the FM state after the SF transition, where both CrSBr sublattices and the Co layer align with the applied field. We also observe MR values of ~300–400% at intermediate field, between field saturated FM state and AFM ground state. These intermediate states are attributed to a gradual, layer-resolved transition from the FM to the AFM configuration. In CrSBr, the exchange length is shorter than the interlayer spacing, resulting in a non-abrupt FM-AFM transition in which magnetic sublattices reverse their orientation sequentially in a layer-by-layer manner[16,24]. Therefore, the FM-AFM transition in the MR curve appears as a sequence of multiple steps. Next, we examine the effect of bias voltage on the measured field-dependent MR. Fig. 2e shows MR traces measured at 11 K for different bias voltages, with the curves are vertically offset for clarity. The characteristic SF-driven MR response of CrSBr persists across all applied biases. However, the MR peak magnitude systematically decreases with increasing bias, as shown in Fig. 2f, and is consistent with previous reports[33,34]. Fig. 2g shows the temperature dependence of MR at a fixed bias of 10 mV. The magnitude of MR decreases with increasing temperature due to thermal suppression of AFM order, eventually disappears above CrSBr transition temperature ($T_N$ ~130 K) (Fig. 2h). A faint residual signal above $T_N$ is attributed to a small MR signal originating from the Co electrode (see the Supplementary Fig. S15). Importantly, once the AFM state is established near zero field, the magnetization of Co electrode can be controlled by applying a small magnetic field, as its coercivity is much smaller than the SF field of CrSBr. We note that in our devices the shape anisotropy of the Co electrode is purposely oriented along the *b*-axis of CrSBr and its magnetization has two stable states, i.e., parallel and antiparallel to the *b*-axis of CrSBr. To highlight this behaviour, the low-field regime of the measured MR response (dotted box in Fig. 2c) is shown in Fig. 2d. For applied bias voltage of +100 mV, we observe that upon switching the magnetization of the Co electrode (green arrow) relative to the Néel vector of CrSBr (purple arrow), we observe a clear two-level MR response, strongly indicative of its utility to detect Néel vector orientation in CrSBr.

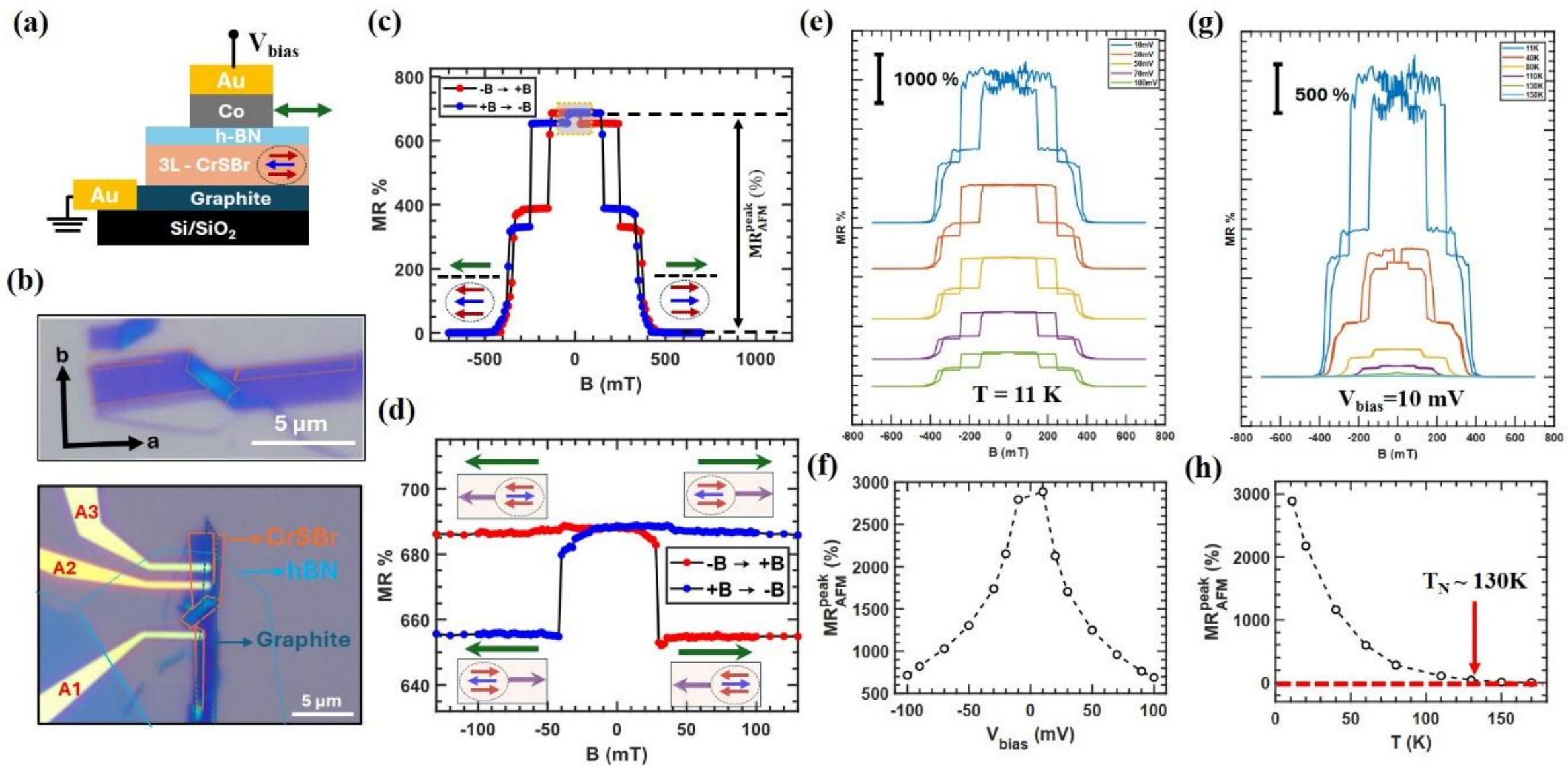

**Fig. 2. Spin-filtering in Co/h-BN/CrSBr heterostructure. (a)** Sideview schematic and measurement geometry of the Co/hBN/CrSBr device. The Co magnetization is indicated by a green arrow, while the AFM ordering in CrSBr is represented by anti-parallel sublattice moments shown in red and blue. **(b)** Top: optical image of the CrSBr flake, with the active region outlined, used in the final device. Bottom: optical image of the fabricated heterostructure device, with individual flakes labeled and outlined. The top contacts A1, A2, and A3 on the hBN barrier consist of 400 nm-wide Co, 450 nm-wide Au, and 450 nm-wide Co electrodes, respectively. **(c)** Vertical MR as a function of magnetic field along easy axis of CrSBr, showing low- and high-resistance states associated with field-induced FM and AFM phases in CrSBr (arrows mark representative states). Forward and backward sweeps are shown in blue and red data, respectively. $\mathbf{MR^{peak}_{AFM}}$ (%) is defined as the difference between the MR at maximum field and that at zero field. **(d)** enlarged view of the low-field region, i.e., boxed region in **(c)**, highlighting spin-valve-like MR behavior due to Co magnetization (green arrow) reversal with respect to a fixed Neel vector (purple arrow). **(e)** MR as a function of magnetic field measured at different bias voltages at T = 11 K. **(f)** MR peak values extracted from **(e)** plotted as a function of bias voltage**. (g)** MR as a function of magnetic field at different temperatures with $V_{bias}$ = 10 mV. **(h)** MR peak values extracted from **(g)** and plotted as a function of bias voltage temperature.

So far, we have demonstrated spin-valve-like behaviour in a vertical Co/hBN/CrSBr tunnel device, wherein orientation of the Co magnetization relative to a fixed AFM configuration in CrSBr produces distinct low- and high-resistance states. This strongly indicate that our approach can resolve the two degenerate Néel vector configurations related by a 180-degree reversal. Previous studies using photoluminescence based on magneto-excitonic coupling, as well as second-harmonic generation spectroscopy exploiting the spatial and temporal symmetries of CrSBr, have shown that the zero-field AFM ground state depends on the polarity of the initial field-polarized state[16,17,24,36]. As illustrated in Fig. 3a, AFM configurations with opposite Néel vector orientations (180-degree reversal) can be realized by initializing the system into a FM state using magnetic fields of opposite polarity. Consequently, once a specific AFM state is stabilized following field saturation in a given direction, the resulting spin-valve response arising from tunnelling between the Co layer and the interfacial CrSBr layer can be used to probe the relative alignment between the Néel vector and the Co magnetization. To this end, we first stabilize the Néel vector $+\widehat{\boldsymbol{N}}$ by driving the system into a FM state applying +1.1 T magnetic field and subsequently reducing the field to zero (Fig. 3b). The Néel vector (purple arrow) and the sublattice moments (red and blue arrows) are illustrated in the inset. The magnetic field is swept within a range below the SF field of CrSBr to preserve a fixed AFM configuration. In field sweep range of 100 mT, the Néel vector in CrSBr remains fixed in one direction, whereas the Co magnetization (green arrow) can be readily manipulated in two distinct directions, i.e., parallel and antiparallel to the Néel vector. As the Co magnetization switches relative to the fixed AFM state, we observe two distinct low- and high-MR states, exhibiting hysteretic behaviour. To confirm the 180° reversal of the Néel vector under opposite field initialization, we repeat the measurement after preparing the AFM state ($-\widehat{\boldsymbol{N}}$) by applying −1.1 T field (Fig. 3c). Remarkably, the low- and high-resistance states reappear at the Co coercive field, but with a hysteresis direction opposite to that observed for the $+\widehat{\boldsymbol{N}}$ initialized AFM state. This inversion reflects the opposite orientation of the Néel vector in the two cases. These observations therefore provide compelling transport evidence that the Néel vector changes sign upon initialization from a ferromagnetic state using opposite field polarity. Next, we perform the same measurement at various bias voltages following −1.1 T initialization (Fig. 3d) and observe a two-state hysteretic behaviour over a range of both positive and negative biases. The MR peaks are extracted by subtracting the forward and backward resistance traces at B=0 T and are plotted as a function of bias voltage in Fig 3e (see Methods for details). Notably, the MR polarity reverses sign near ∼±5 mV. This behaviour can be qualitatively understood within the framework of the Jullière tunnelling model[42], where spin is assumed to be conserved during tunnelling between majority and minority spin channels. In this model, the magnitude and sign of the MR are governed by the spin polarization of the electrodes, which is determined by the relative density of states (DOS) of spin-up (↑) and spin-down (↓) carriers at the Fermi level, and the MR is defined as:

$$MR = \frac{R_{AP} - R_P}{R_P} = \frac{2P_{Co}P_{CrSBr}}{1 - P_{Co}P_{CrSBr}}$$

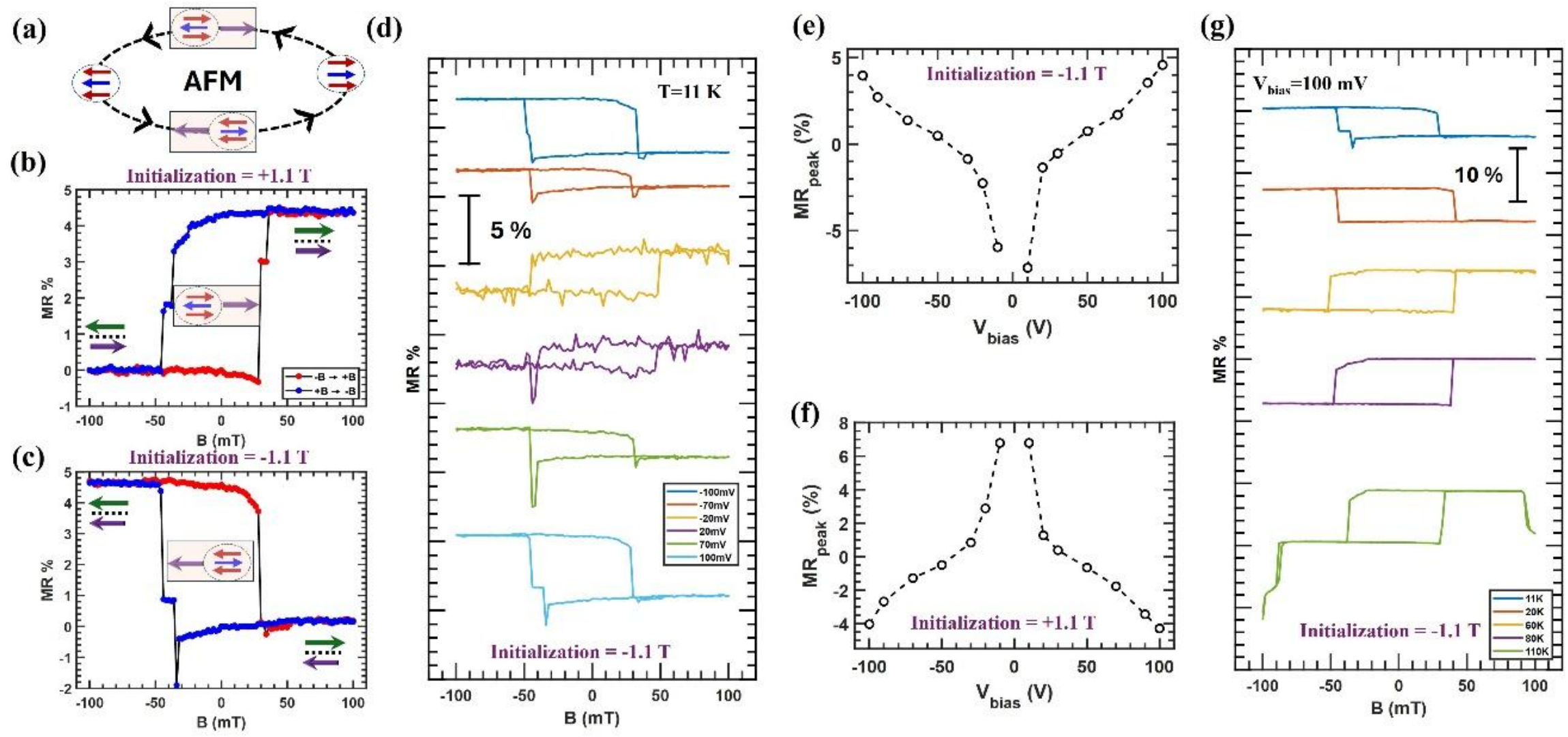

**Fig. 3. Tunnel-transport detection of Néel vector reversal in CrSBr**. **(a)** Schematic showing two antiparallel (180°) Néel vector states stabilized by opposite high-field initialization conditions. **(b,c)** Field-dependent MR measured below the SF transition following initialization by +1.1 T (b) and −1.1 T (c), at T=11 K and $V_{bias}$=100 mV. The opposite hysteresis polarity reflects the reversal of the Néel vector through spin-dependent tunneling between Co and CrSBr. The AFM sublattice moments are shown by red and blue arrows, their corresponding Néel vector by the purple arrow, and the Co magnetization by the green arrow. **(d)** Bias-dependent MR hysteresis measured below the SF field following −1.1 T initialization at T = 11 K. **(e)** MR peak values extracted from the bias-dependent hysteresis measurements shown in (c). **(f)** MR peak values as a function of bias voltage after +1.1 T initialization, confirming opposite polarity of hysteresis as compared to (d). **(g)** MR hysteresis measured at various temperatures at $V_{bias}$ = 100 mV following −1.1 T initialization.

Here, $R_{AP\ (P)}$ denotes the resistance in the antiparallel (parallel) magnetic configuration, while $P_{Co\ (CrSBr)}$ represent the spin polarization of Co and CrSBr, respectively. The spin polarization of each electrode is determined by the relative density of states (DOS) of majority (↑) and minority (↓) spin carriers at the Fermi level, given by:

$$P = \frac{DOS_{\uparrow} - DOS_{\downarrow}}{DOS_{\uparrow} + DOS_{\downarrow}}$$

Previous density functional theory studies have shown that the electronic band structure of monolayer CrSBr is strongly spin dependent[43-45]. Accordingly, the polarity reversal of the MR observed in Fig. 3d-e can be understood in terms of a bias-induced shift of the Fermi level across spin-polarized bands of monolayer CrSBr, such that the condition $DOS_{\uparrow} < DOS_{\downarrow}$ is realized. This redistribution of the spin-resolved DOS changes the effective spin polarization of the tunneling channel, ultimately leading to the observed sign reversal of the MR. Furthermore, since opposite field-initialization conditions reverse the sign of the Néel vector, the corresponding low-field MR hysteresis also reverses polarity. Consistent with this picture, the MR measured as a function of bias voltage following a +1.1 T initialization field (Fig. 3f) exhibits an exactly opposite trend compared to the case of −1.1 T initialization. Our experimental results demonstrate that this approach for detecting the Néel vector orientation is robust across a wide range of bias voltages and consistently exhibits opposite hysteresis polarities for opposite Néel vector orientations. We further investigate the temperature dependence of the tunneling response. Fig. 3g shows measured MR during magnetic field sweeps between ±100 mT at a bias voltage of 100 mV, following initialization with a −1.1 T magnetic field, at various temperatures. The observed hysteresis, reflecting the relative orientation of the Co magnetization and the CrSBr AFM state, remains clearly visible up to 110 K, as the SF field of CrSBr continues to exceed the coercivity of the Co layer in this temperature range. In addition, by engineering a lower coercivity in the Co electrode using a wider electrode (see Supplementary Fig. S16), the detection of the Néel vector can be sustained up to the Néel temperature of CrSBr.

Having clearly established that spin-dependent tunnelling between Co and CrSBr across an ultrathin tunnel barrier enables electrical detection of the Néel vector reversal in CrSBr, we next verify that this effect specifically originates from the ferromagnetic spin polarizer electrode and examine its reproducibility across different devices and tunnel barriers. To this end, we first measure an additional contact on Device A consisting of non-magnetic (gold, Au) electrode (Device A2), i.e., Co is replaced with a non-magnetic (Au) electrode in our measurements. A side-view schematic of the device is shown in the inset of Fig. 4a, while the corresponding optical image highlighting contact A2 is presented in Fig. 2b. The MR measured over the full magnetic field range exhibits the characteristic response of CrSBr, with distinct high- and low-MR states corresponding to the AFM and FM phases across the SF transition, respectively (Supplementary Fig. S2, S3). To verify that the low-field hysteresis observed in previous measurements originates from the Co magnetization acting as a probe of the CrSBr Néel vector, we perform magnetic field sweeps within a field range smaller than the CrSBr SF field following initialization at −1.1 T (Fig. 4a). As expected, in the absence of Co spin-polarized electrode, no measurable hysteresis is observed. This result confirms that the low-field hysteretic response is unique to spin-dependent tunnelling between FM layer and CrSBr.

Furthermore, we observe similar behaviour in a device wherein Co and CrSBr are separated by trilayer graphene, instead of hBN tunnel barrier. The inset of Fig. 4b shows a side-view schematic of the Co/Gr/CrSBr heterostructure (Device B), where the Co electrode width is 800 nm (see Supplementary Fig. S8-S10 for additional details). Magnetic field sweeps performed over a reduced field range (±50 mT) reveal a clear hysteretic response associated with reversal of the Co magnetization (Fig. 4b). Reversing the Néel vector initialization polarity from −1.1 T to +1.1 T produces an opposite hysteresis polarity, as shown in Fig. 4c. The corresponding coercive fields are approximately 25 mT, smaller than those observed for Device A1 (~42 mT) and Device A3 (~38 mT) (Supplementary Information S16). The reduced coercivity for the wider Co contact suggests that magnetization reversal increasingly proceeds through domain-wall nucleation and propagation, rather than coherent single-domain switching, thereby lowering the switching field.

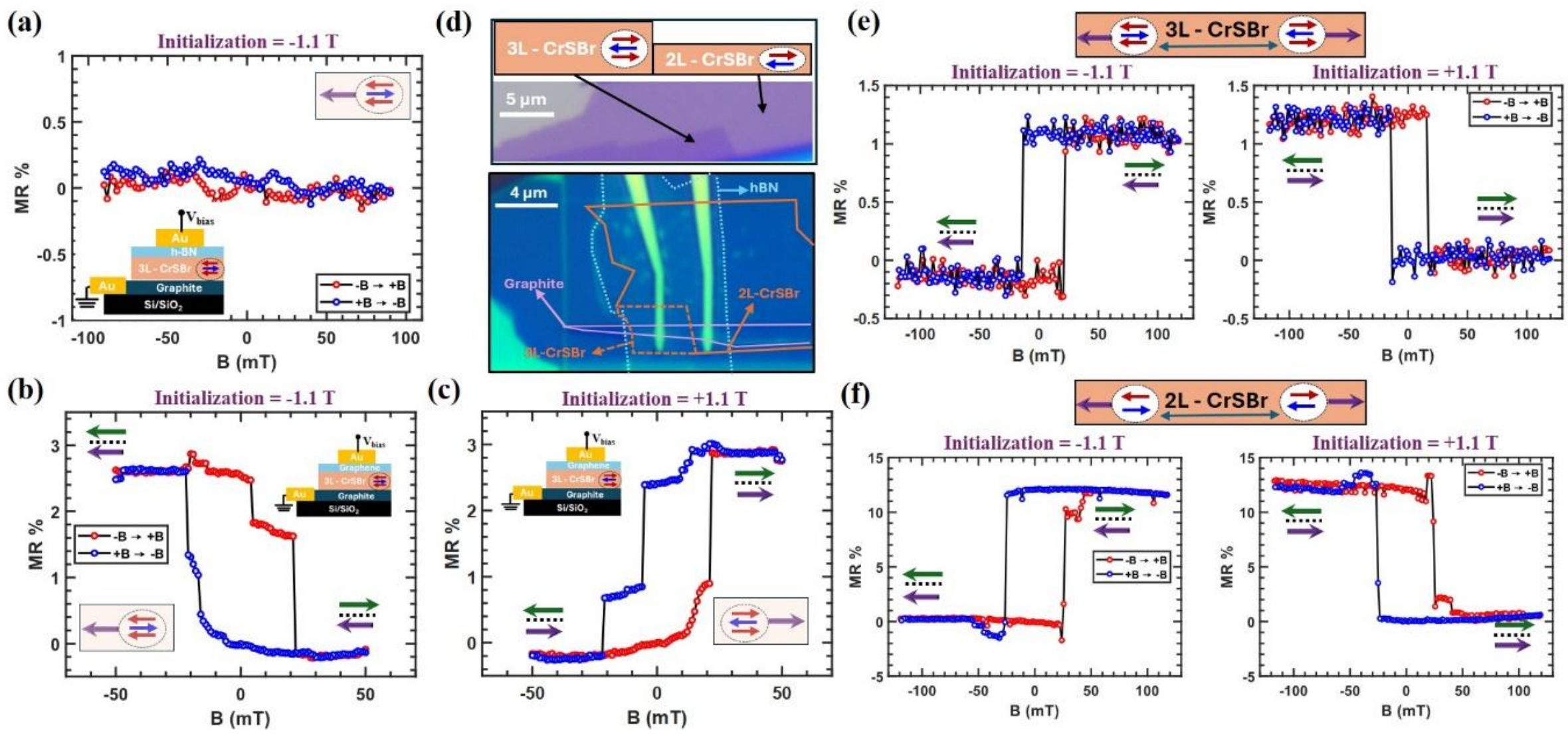

**Fig. 4. Verification of Co-induced spin filtering and layer-resolved response in CrSBr. (a)** MR response of an Au/hBN/CrSBr device A2 (sideview schematic shown in inset), between SF fields after initialization of the Néel state by -1.1 T, showing absence of hysteresis. **(b, c)** MR hysteresis measured between the SF fields in a Co/graphene/CrSBr device (Device B) after stabilization of the Néel state using magnetic fields of (b) −1.1 T and (c) +1.1 T. Inset: side-view schematic of the heterostructure and the corresponding magnetic configurations associated with the MR states. **(d)** CrSBr flake with even (2L) and odd (3L) layer regions outlined in the top panel, and the corresponding Co/hBN/CrSBr device (Device C) fabricated from the same flake shown in the bottom panel. The Co electrodes in device C are aligned with the 2L and 3L regions of CrSBr, respectively, and all flakes in the final device are outlined and labeled. **(e)** MR hysteresis in a Co/hBN/3L-CrSBr device following Néel-state initialization at −1.1 T (left) and +1.1 T (right). The associated magnetic configurations are depicted alongside each MR state. **(f)** MR hysteresis obtained from a Co/hBN/2L-CrSBr device after Néel-state initialization at −1.1 T (left) and +1.1 T (right). The magnetic configurations associated with each MR state are illustrated.

Finally, to verify that the observed hysteresis is not unique to odd-layer CrSBr, which possesses an uncompensated sublattice moment, we investigate Device C containing both even- and odd-layer CrSBr regions. The top panel of Fig. 4d shows optical images of the 2L and 3L terraces of the CrSBr flake used for device fabrication, while the bottom panel shows the final fabricated device with contacts C1 and C2 positioned on the 3L (Device C1) and 2L (Device C2) regions, respectively. We first measure Device C1 and observe a clear hysteretic response consistent with spin filtering between Co and 3L CrSBr. As shown in Fig. 4e, low- and high-resistance states emerge depending on the relative orientation between the Co magnetization and the CrSBr Néel vector after initialization at −1.1 T (left panel) and +1.1 T (right panel). Remarkably, Device C2, comprising Co/hBN on 2L CrSBr, exhibits a similar hysteretic response following opposite initialization fields (Fig. 4f). The observation of hysteresis in 2L CrSBr, where the magnetic moments of adjacent layers are fully compensated, confirms that the distinct MR states in Co/hBN/CrSBr tunnel devices originate from spin filtering at the interfacial CrSBr layer adjacent to the Co/hBN contact. Since CrSBr is an A-type AFM with ferromagnetically ordered individual layers, each CrSBr monolayer possesses a spin-polarized electronic band structure despite the overall compensated AFM state. Consequently, the tunnelling resistance is governed by the relative alignment between the Co magnetization and the spin polarization of the interfacial CrSBr layer. More broadly, these results establish a general mechanism for electrical Néel vector detection in layered A-type AFMs with spin-polarized electronic bands.

By employing a vertical device configuration, wherein varying thickness of atomically thin CrSBr layers are integrated with a FM electrode through a h-BN tunnel barrier, we experimentally demonstrate electrical readout of 180-degree reversal of Néel vector. We note that this strategy is generalizable to other A-type AFMs as well. For instance, for out-of-plane easy-axis systems, such as $CrPS_4$, a FM electrode with perpendicular magnetic anisotropy, can be integrated across a tunnel barrier to achieve similar spin-sensitive readout of the Néel vector. Combining with the ability of emergent vdW materials to realize efficient and unconventional spin-current for electrical control of magnetic order, our work which forms the foundation for demonstration of AFM-based ultrafast magnetic memory and other spintronics devices.

## Methods

**Device fabrication.** High-quality CrSBr single crystals used in this study were synthesized via a direct solid-vapor reaction in a box furnace following previously reported procedures[18]. hexagonal boron nitride (hBN) crystals were prepared using established growth methods[46]. Mechanical exfoliation of CrSBr, hBN, and graphite flakes was performed inside an Ar-filled glovebox onto separate Si/$SiO_2$ substrates with a 300 nm oxide layer. For tunnel barrier layer, thin hBN flakes were exfoliated onto Si/$SiO_2$ substrates with a 90 nm oxide thickness. Suitable flakes were first identified by optical microscopy and subsequently characterized by atomic force microscopy to accurately determine their thickness and layer number. Exfoliated CrSBr flakes exhibit a characteristic rectangular shape, with the crystallographic *a*-axis aligned along the long edge of the flake.

vdW heterostructures comprising Co/hBN/CrSBr/graphite and Co/graphene/CrSBr/graphite stacks were assembled on separate thermally oxidized Si/$SiO_2$ substrates using a custom-built dry-transfer system. The transfer process employed a polycarbonate (PC) film supported on a polydimethylsiloxane (PDMS) stamp for pickup and release of the constituent two-dimensional heterostructures. Following heterostructure assembly, electrical contacts were defined by electron-beam lithography using a PMMA/MMA bilayer resist, followed by electron-beam evaporation of Co (25 nm)/Au (5 nm) or Cr (5 nm)/Au (25 nm) (in Device A2) electrodes. The Co electrodes had widths of 400 nm (Device A1), 450 nm (Devices A2, A3, C1, and C2), and 800 nm (Device B). Subsequently, Cr (5 nm)/Au (55 nm) electrodes were deposited to form large-area contact pads for wire bonding and low-temperature electrical transport measurements.

**Electrical measurements.** Electrical transport measurements were performed under high-vacuum conditions ($<10^{-5}$ mTorr) over a variable temperature range. An electromagnet mounted on a rotational stage was used to apply the magnetic field in the plane of the device, specifically along the crystallographic b-axis of CrSBr. A dc bias voltage was applied using Keithley 2400 SourceMeter and Keithley 2450 SourceMeter units, while the resulting current was recorded simultaneously to determine the device resistance.

For magnetic field sweeps spanning the full field range across the spin-flip transition, the magnetoresistance was defined as MR (%) $= \frac{R(B)-R(FM)}{R(FM)} \times 100$, where R(B) is the resistance measured at magnetic field B, and $R_{FM}$ is the average resistance in the fully field-polarized ferromagnetic (FM) state at maximum positive and negative magnetic fields. The peak value of MR ($MR_{AFM}^{peak}$ (%)), was extracted from the difference between the MR at B=0 and that in the maximum field induced FM state. For small magnetic field sweep ranges probing the low-field hysteresis, the MR was defined as MR (%) $= \frac{R-R(low)}{R(low)} \times 100$, where R(low) corresponds to the low-resistance state at B=0. In this case, the peak hysteretic magnetoresistance, $MR_{Peak}$, was obtained from the difference between the forward- and backward-field sweep MR values at B=0.

## Acknowledgments

S.S. acknowledges the financial support from U.S. Office of Naval Research (ONR) under Award No. N00014-23-1-2751; National Science Foundation (NSF) through Grant No. ECCS-2531211; NSF-CAREER Award through Grant No. ECCS-2339723; and W.M. Keck Foundation. J.K. acknowledges the financial support from U.S. Department of Energy, Office of Science, Office of Basic Sciences, through award No. DE-SC002549 (for measurements and device fabrication); from ONR under Award No. N00014-23-1-2751; and from NSF-CAREER Award under Grant No. DMR-2339309. The work at University of Texas at Dallas is supported by US Air Force Office of Scientific Research (AFOSR) (FA9550-19-1-0037), National Science Foundation (NSF) (DMREF-2324033 and 2516364) and Office of Naval Research (ONR) (N00014-23-1-2020). J.H.E. acknowledges the support for hBN crystal growth from the U. S. Office of Naval Research under award number N00014-26-1-2111. K.W. and T.T. acknowledge support from the JSPS KAKENHI (Grant Numbers 21H05233 and 23H02052), the CREST (JPMJCR24A5), JST and World Premier International Research Center Initiative (WPI), MEXT, Japan.

## Author Contributions

S.S. and J.K. designed the experiments and supervised the research. R.P. and R.K.B. prepared the devices with assistance from A.S., P.S., Z.C., A.T. and I.K. R.P. performed the measurements and analyzed data. W.L., Z.Z. and B.L. grew the bulk crystals of CrSBr. T.P., J. H. E., K.W., and T.T. provided the bulk hBN crystals. All authors contributed to write the manuscript.

## Competing interests

The authors declare no competing interests.

# Supplementary Information

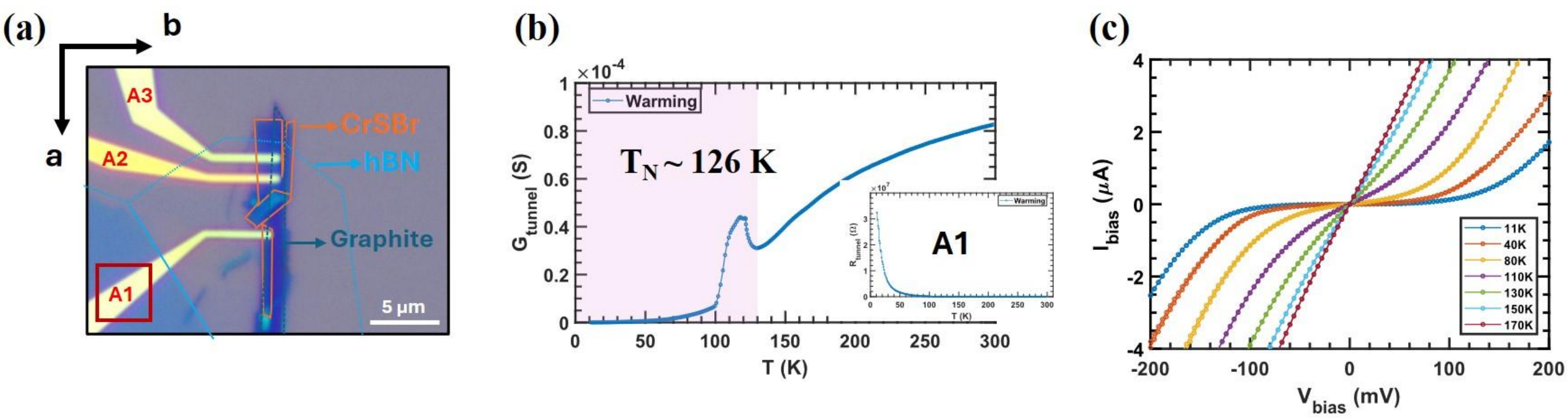

**Fig. S1. Electrical characterization of Au/hBN/CrSBr Device A1: (a)** Optical image of the fabricated Au/hBN/CrSBr tunnel junction device (Device A1). Individual flakes are outlined and labelled, and contact A1 is highlighted by a box. **(b)** Device conductivity as a function of temperature, showing a local conductivity change near ~126 K, close to the Néel temperature of CrSBr. Inset: Resistance versus temperature, exhibiting an increase in resistance upon cooling, consistent with the semiconducting nature of CrSBr. **(c)** Nonlinear current–voltage (I-V) characteristics measured at different temperatures.

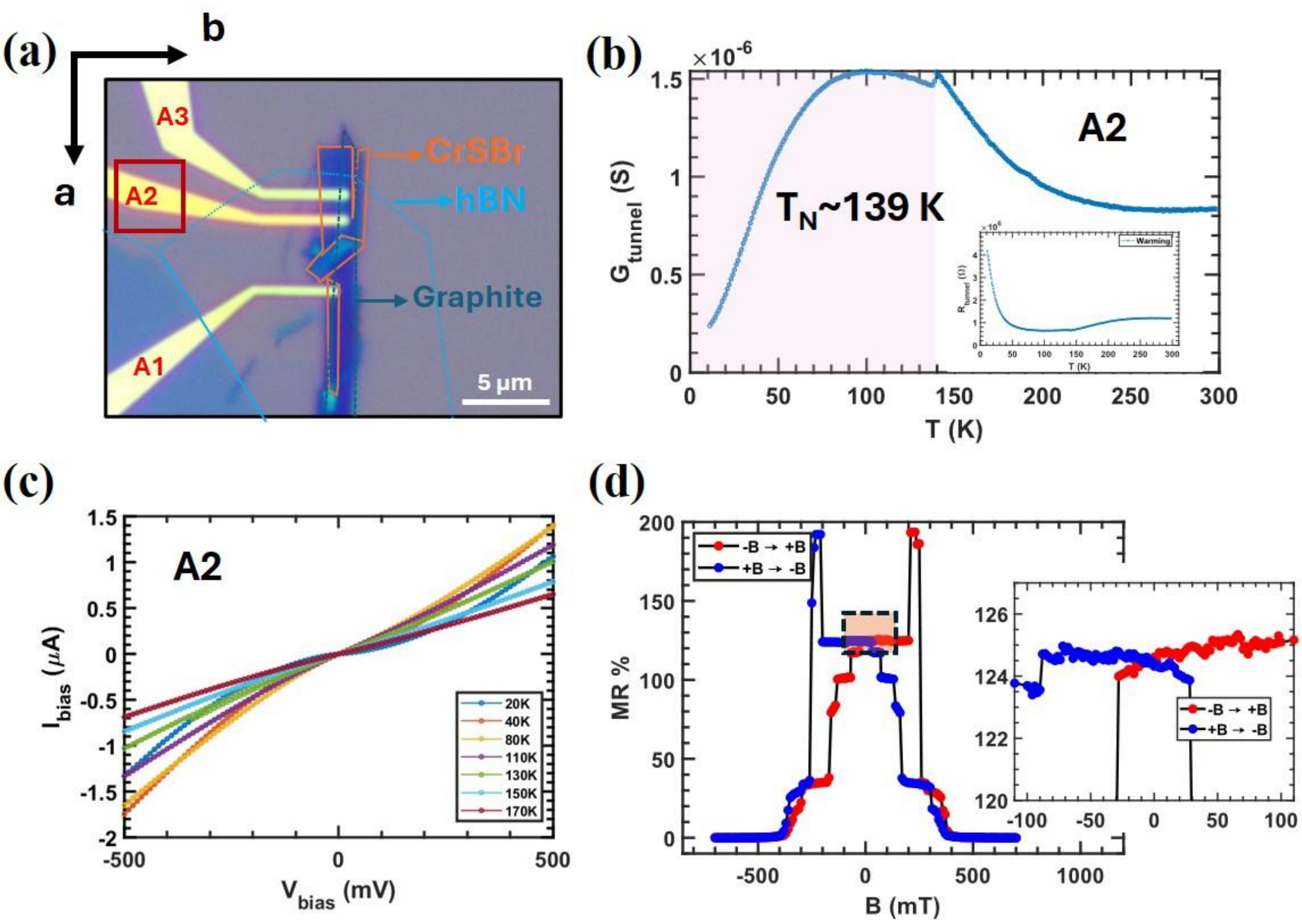

**Fig. S2. Electrical characterization of Au/hBN/CrSBr Device A2: (a)** Optical image of the fabricated Au/hBN/CrSBr tunnel junction device (Device A2). Individual flakes are outlined and labelled, and contact A2 is highlighted by a box. **(b)** Device conductivity as a function of temperature, showing a local conductivity maximum near ~139 K, close to the Néel temperature of CrSBr. Inset: Resistance versus temperature, exhibiting an increase in resistance upon cooling, consistent with the semiconducting nature of CrSBr. **(c)** Nonlinear I-V characteristics measured at different temperatures. **(d)** Representative of magnetoresistance (MR) as a function of magnetic field, showing a sharp spin-flip (SF) transition separating the AFM state from the high-field FM state. Inset: Enlarged low-field region showing that the forward (backward) field sweep does not exhibit any switching in the positive (negative) field range within ±100 mT, indicating the absence of spin-valve-like behaviour in the absence of a FM electrode.

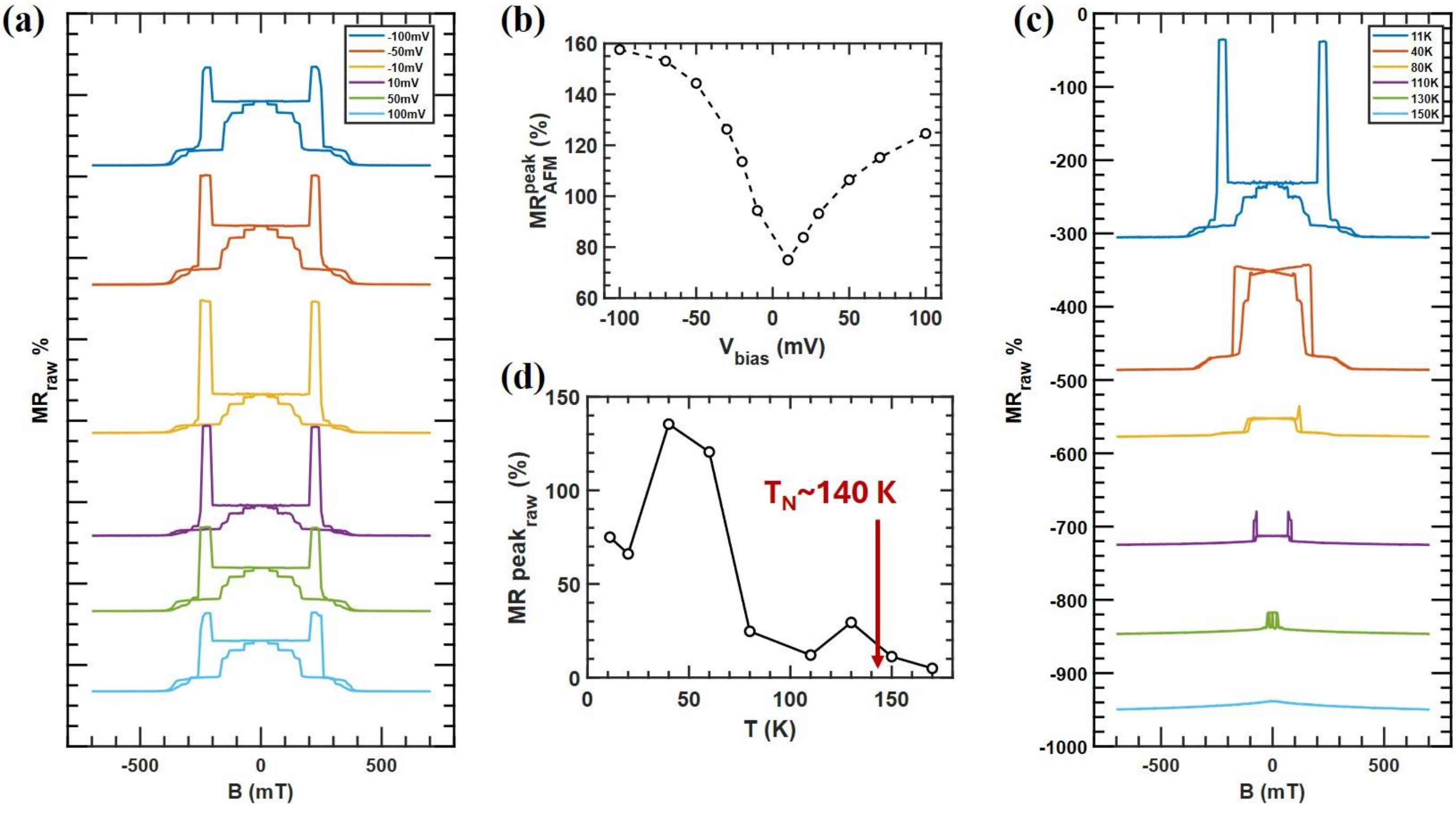

**Fig. S3. Bias- and temperature-dependent magnetoresistance of Au/hBN/CrSBr Device A2 over the full magnetic field range. (a)** MR as a function of magnetic field (B) measured at different bias voltages. The peak features near the SF transition are suppressed with increasing bias magnitude, suggesting that these peaks originate from domain pinning effects. Their reduction at higher bias is likely caused by enhanced bias current and associated local Joule heating. **(b)** MR peak amplitudes extracted from panel (a) as a function of bias voltage. **(c)** Temperature-dependent MR versus B measurements. The sharp MR peaks near the SF transition gradually disappear with increasing temperature, consistent with domain pinning effects being more pronounced at low temperatures. **(d)** MR peak amplitudes extracted from panel (c) as a function of temperature, showing that the MR signal vanishes above the Néel temperature, $T_N \sim 140$ K.

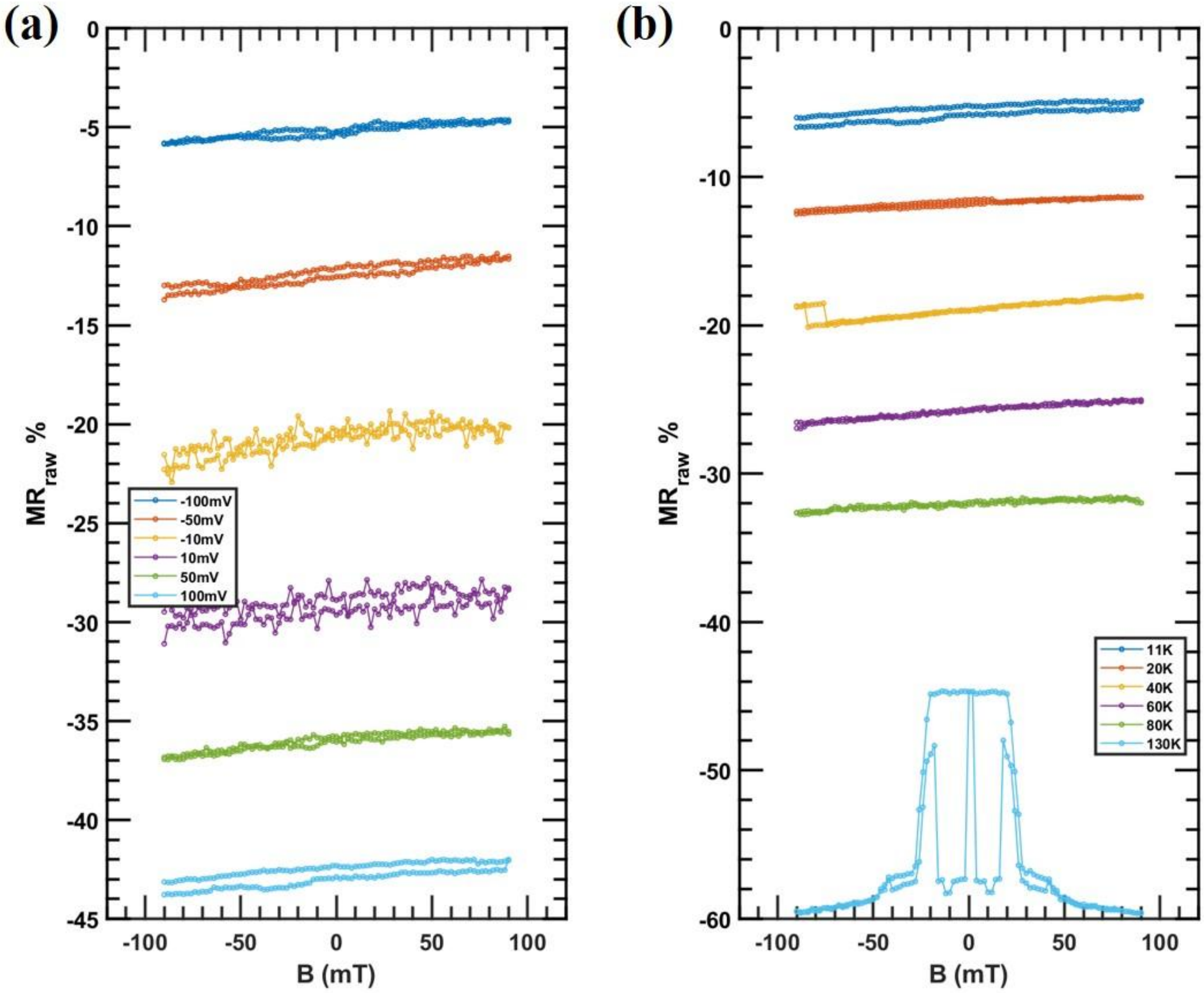

**Fig. S4: MR within the SF transition of Device A2: (a)** MR as a function of B field measured within the SF transition region at different bias voltages. The Néel state is initialized by applying a −1.1 T magnetic field. The absence of hysteretic behaviour in this measurement shows that hysteresis observed in Co-based devices originates from spin-dependent tunnelling between Co and CrSBr. **(b)** MR as a function of B within the SF transition region measured at different temperatures. No hysteresis is observed at any temperature; however, near ∼130 K, the SF transition enters the measured field window, resulting in the appearance of the SF transition feature.

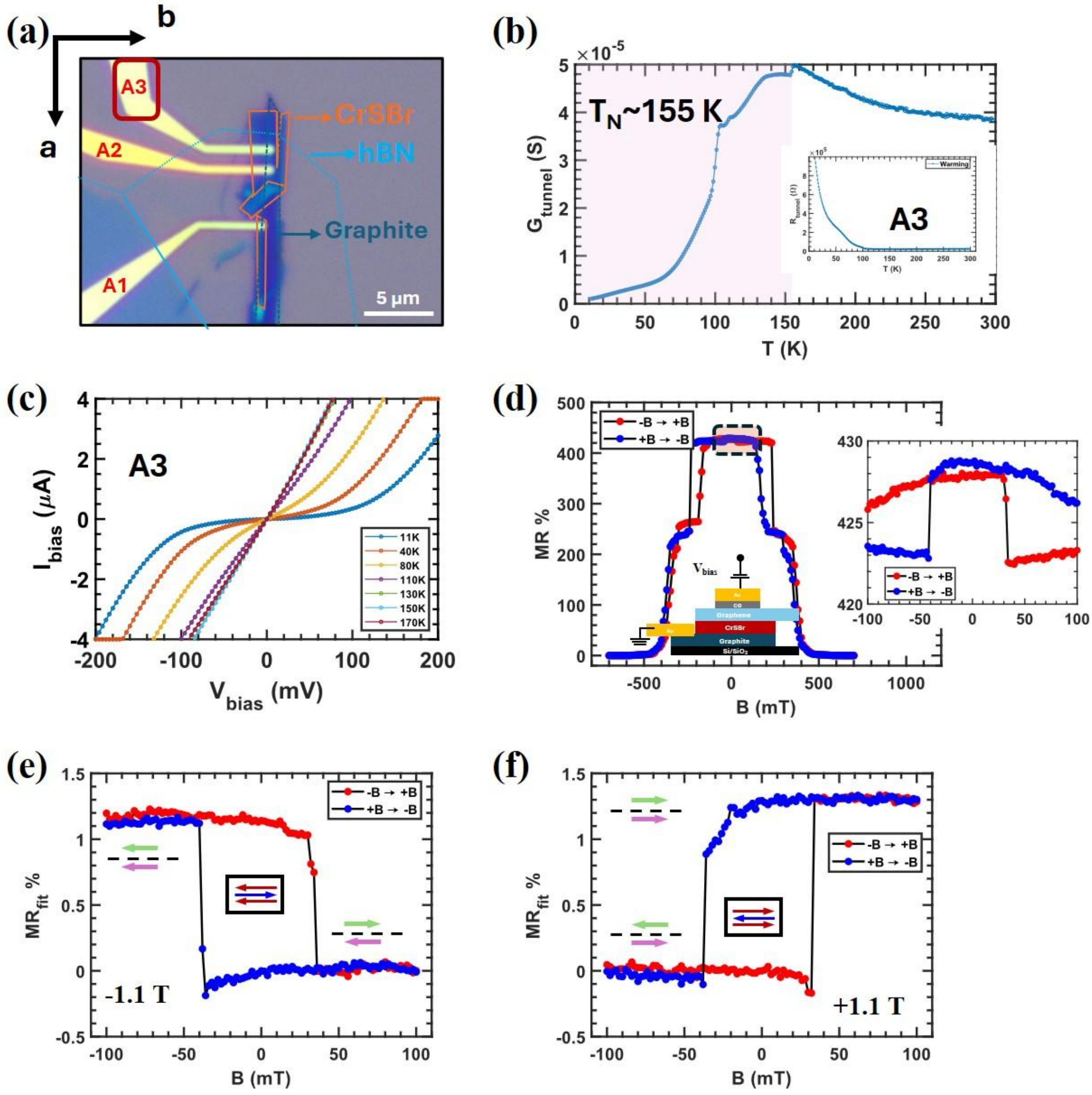

**Fig. S5. Electrical characterization of Co/hBN/CrSBr Device A3: (a)** Optical image of the fabricated Co/hBN/CrSBr tunnel junction device (Device A3). Contact A3 is highlighted by a box. **(b)** Conductivity (G) as a function of T, showing a local conductivity maximum near ∼155 K, close to the Néel temperature of CrSBr. Inset: Resistance versus temperature, showing an increase in resistance as temperature decreases, consistent with the semiconducting behaviour of CrSBr. **(c)** Nonlinear I-V characteristics measured at different temperatures. **(d)** Representative MR as a function of magnetic field, showing a SF transition separating the AFM state from the FM state. Inset: Enlarged low-field region showing that the switching appears in the negative (positive) field range for the forward (backward) field sweep within ±100 mT, demonstrating spin-valve-like behaviour induced by the presence of the Co electrode. **(e,f)** Hysteretic MR response measured within a 100 mT field sweep range after initializing the magnetic state with (e) −1.1 T and (f) +1.1 T. The observed two-state MR behavior confirms that the MR response originates from the relative orientation between the CrSBr Néel vector (purple arrow) and the Co magnetization (green arrow).

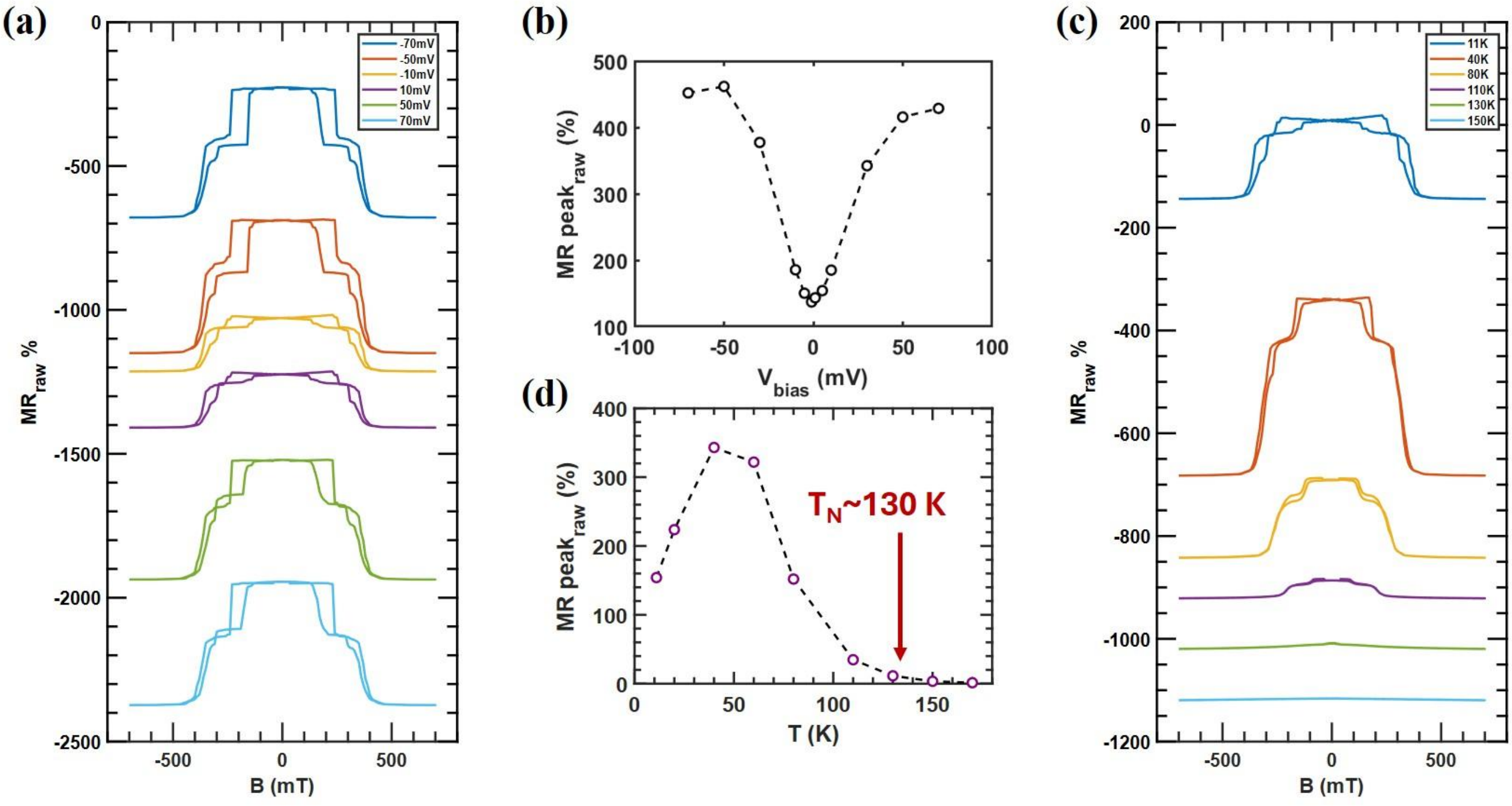

**Fig. S6.: Bias voltage and temperature dependence of MR in complete field sweep range for Device A3. (a)** MR as a function of B measured at different bias voltages. **(b)** MR peak amplitudes extracted from panel (a) as a function of bias voltage. **(c)** Temperature-dependent MR vs. B measurements. **(d)** MR peak amplitudes extracted from panel (c) as a function of temperature, showing that the MR signal vanishes above the Néel temperature, $T_N \sim 130$ K.

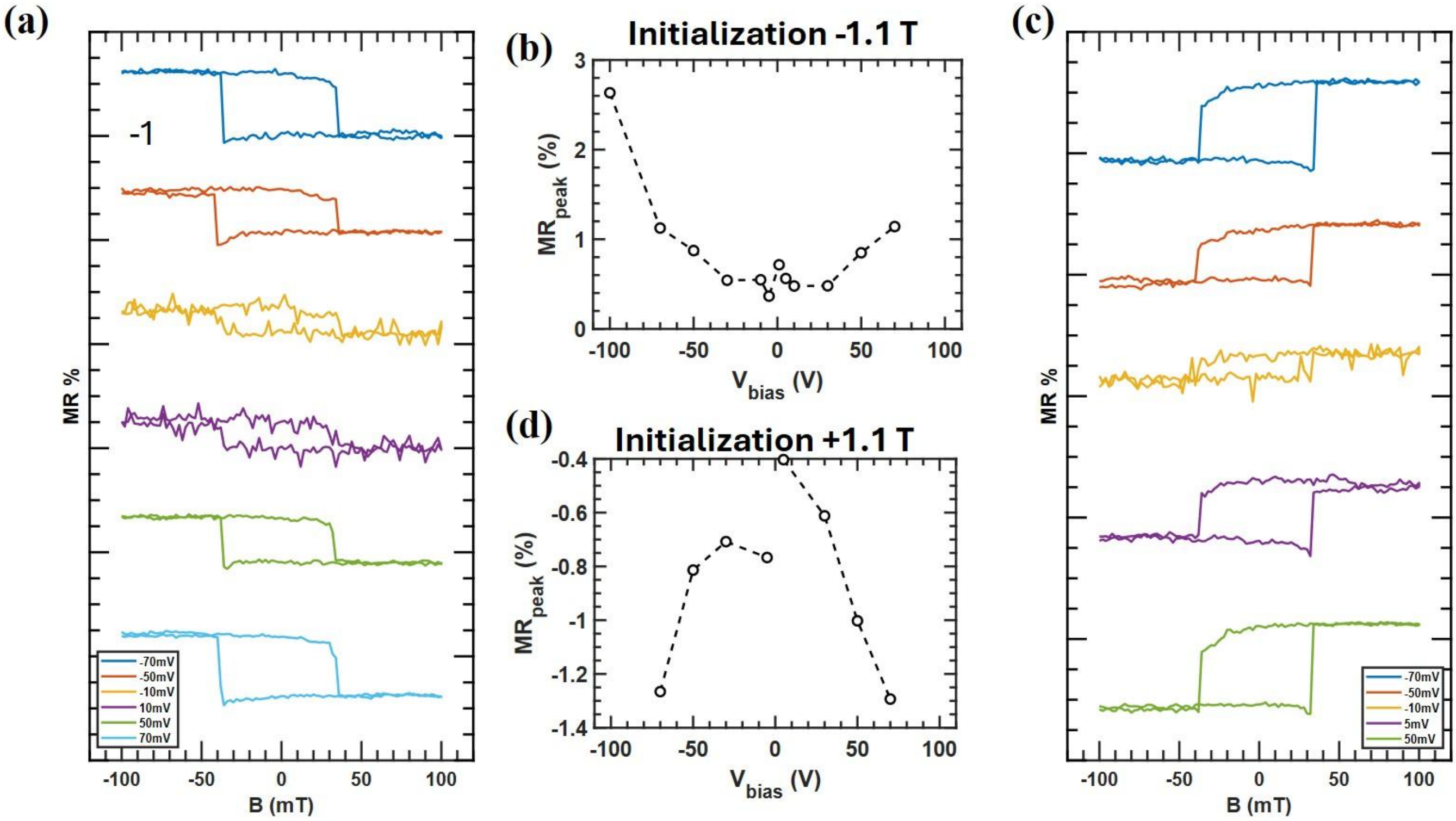

**Fig. S7: MR within the SF transition of Device A3: (a)** MR as a function of B within the SF transition region measured at different bias voltages. The Néel state is initialized by applying a −1.1 T magnetic field. The observed two-state MR arises from spin-dependent tunnelling between Co and CrSBr. **(b)** Extracted MR difference at B=0 T (MR peak) as a function of bias voltage. **(c)** MR as a function of B measured after −1.1 T field initialization, showing an opposite sign of hysteretic MR compared to panel (a). **(d)** Extracted MR peak values from panel (c) as a function of bias voltage.

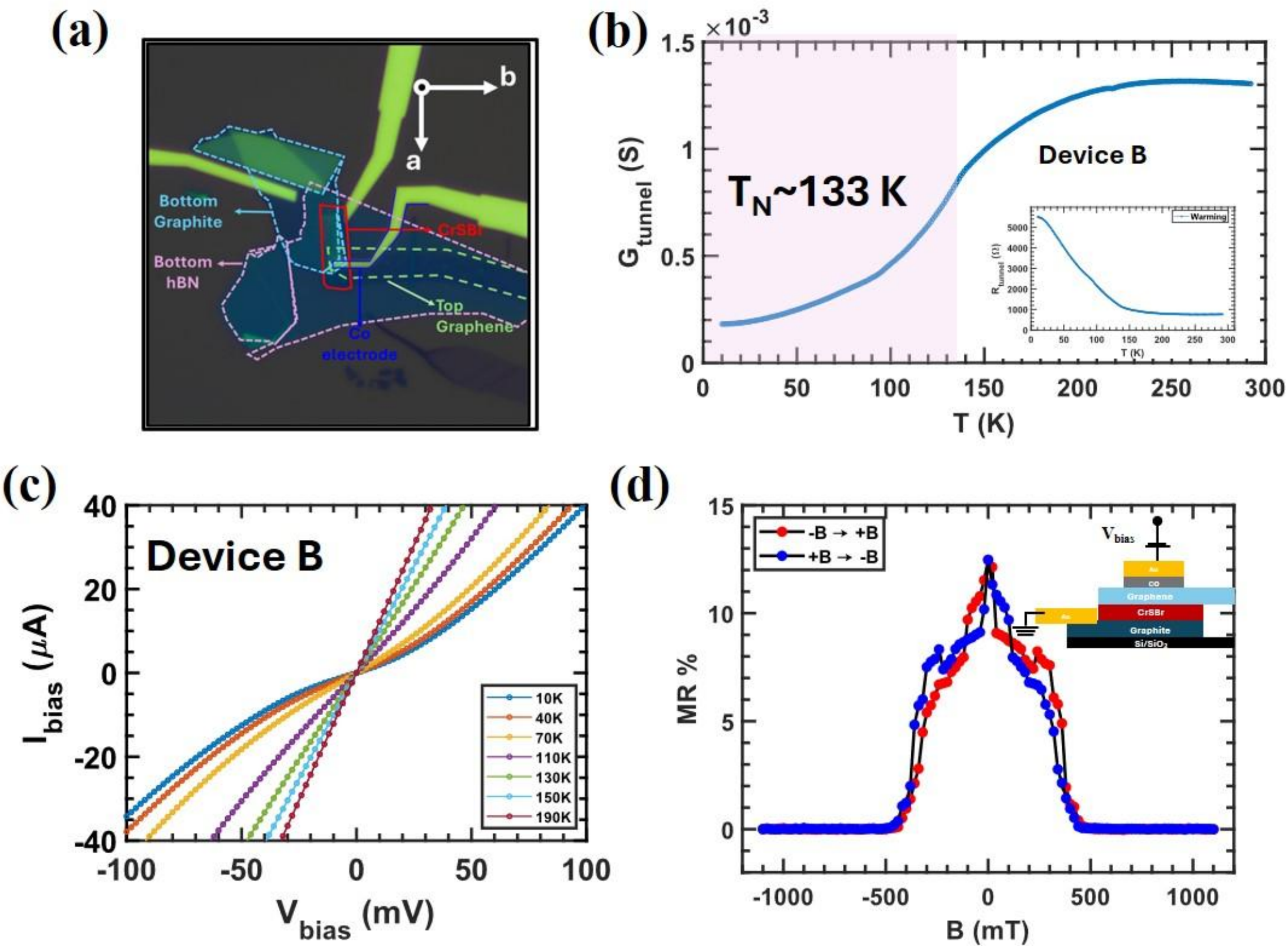

**Fig. S8. Electrical characterization of Co/graphene/CrSBr Device B:** (a) Optical image of the fabricated Co/graphene (3L)/CrSBr (3L) device (Device B). Individual flakes are outlined and labelled. The Co electrode width is 800 nm. **(b)** Conductivity as a function of temperature (T), showing a local conductivity maximum near ∼133 K, close to the Néel temperature of CrSBr. Inset: R vs. T showing semiconducting behaviour of CrSBr; however, the absolute resistance is approximately an order of magnitude lower than that of hBN-based barrier devices**.** **(c)** Nonlinear I-V characteristics measured at different temperatures. **(d)** Representative magnetoresistance (MR) as a function of magnetic field, where the MR peak near zero field arises from spin-dependent transport between Co and CrSBr.

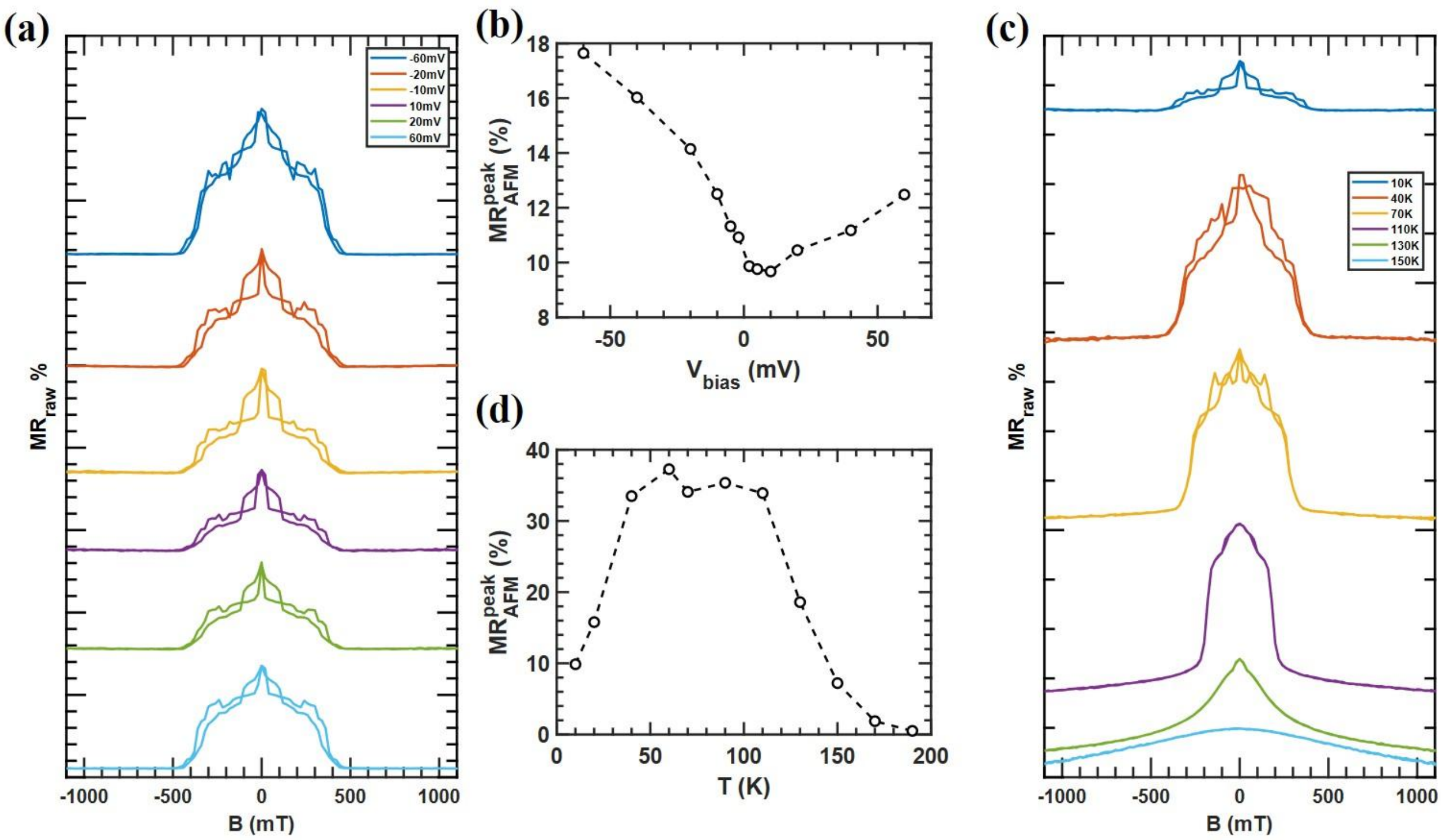

**Fig. S9.: Bias- and temperature-dependent magnetoresistance of Co/graphene/CrSBr Device B over the full magnetic field range. (a)** MR as a function of B measured at different bias voltages. **(b)** MR peak amplitudes extracted from panel (a) as a function of bias voltage. **(c)** Temperature-dependent MR vs. B measurements, showing that the SF transition signal disappears above the Néel temperature, $T_N \sim 130$ K. **(d)** MR peak amplitudes extracted from panel (c) as a function of temperature. The measurable MR above $T_N$ may originate from the Co electrode response; however, no SF transition is observed above 130 K.

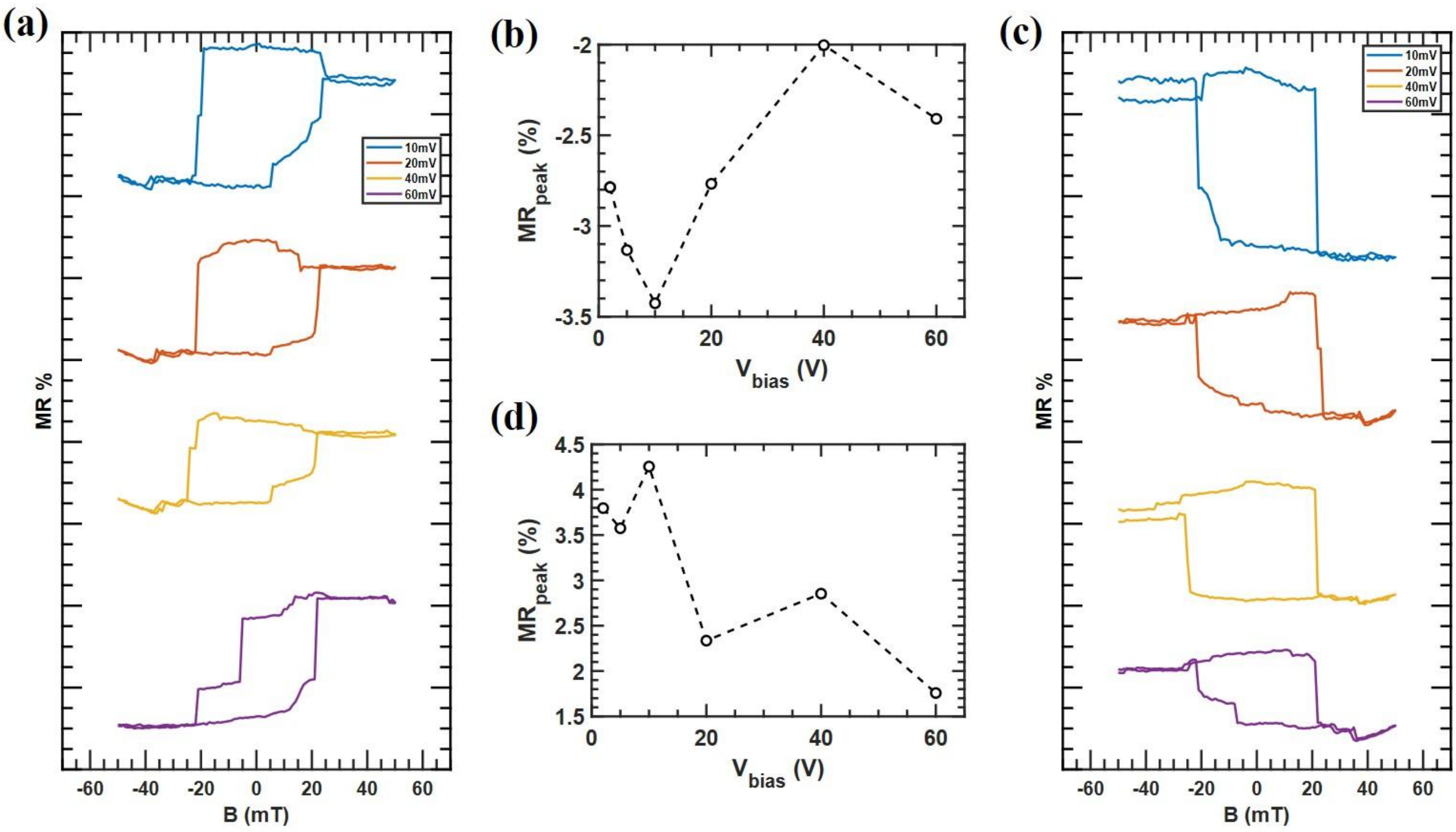

**Fig. S10: MR within the SF transition of Device B: (a)** MR as a function of B within the SF transition region measured at different bias voltages. The Néel state is initialized by applying a −1.1 T magnetic field. The observed hysteresis in MR arises from spin-dependent transport between Co and CrSBr. (b) Extracted MR peak as a function of bias voltage. (c) MR as a function of B measured after +1.1 T field initialization of the Néel state, showing an opposite sign of hysteresis compared to panel (a). (d) Extracted MR peak values from panel (c) as a function of bias voltage.

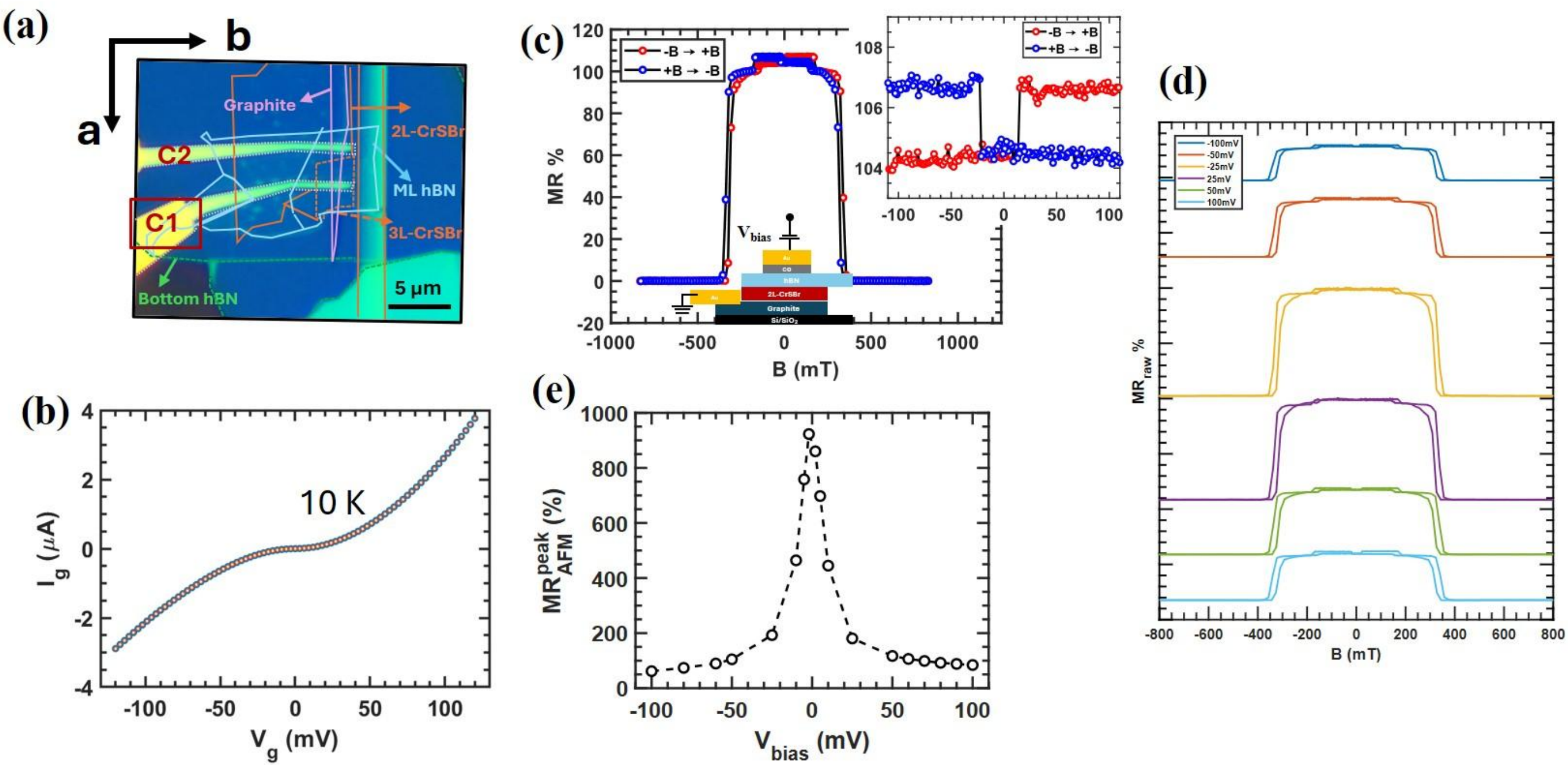

**Fig. S11. Electrical characterization of Co/hBN/3L-CrSBr Device C1: (a)** Optical image of the fabricated Co/hBN/2L-CrSBr tunnel junction device (Device C1). Individual flakes are outlined and labelled. **(b)** Nonlinear I-V characteristics measured at T=10 K. **(c)** Representative MR as a function of B. Inset: Spin-valve-like behaviour associated with reversal of Co magnetization under applied B. **(d)** MR as a function of B over the full field sweep range at different bias voltages. **(e)** MR peak values extracted from panel (d) as a function of bias voltage.

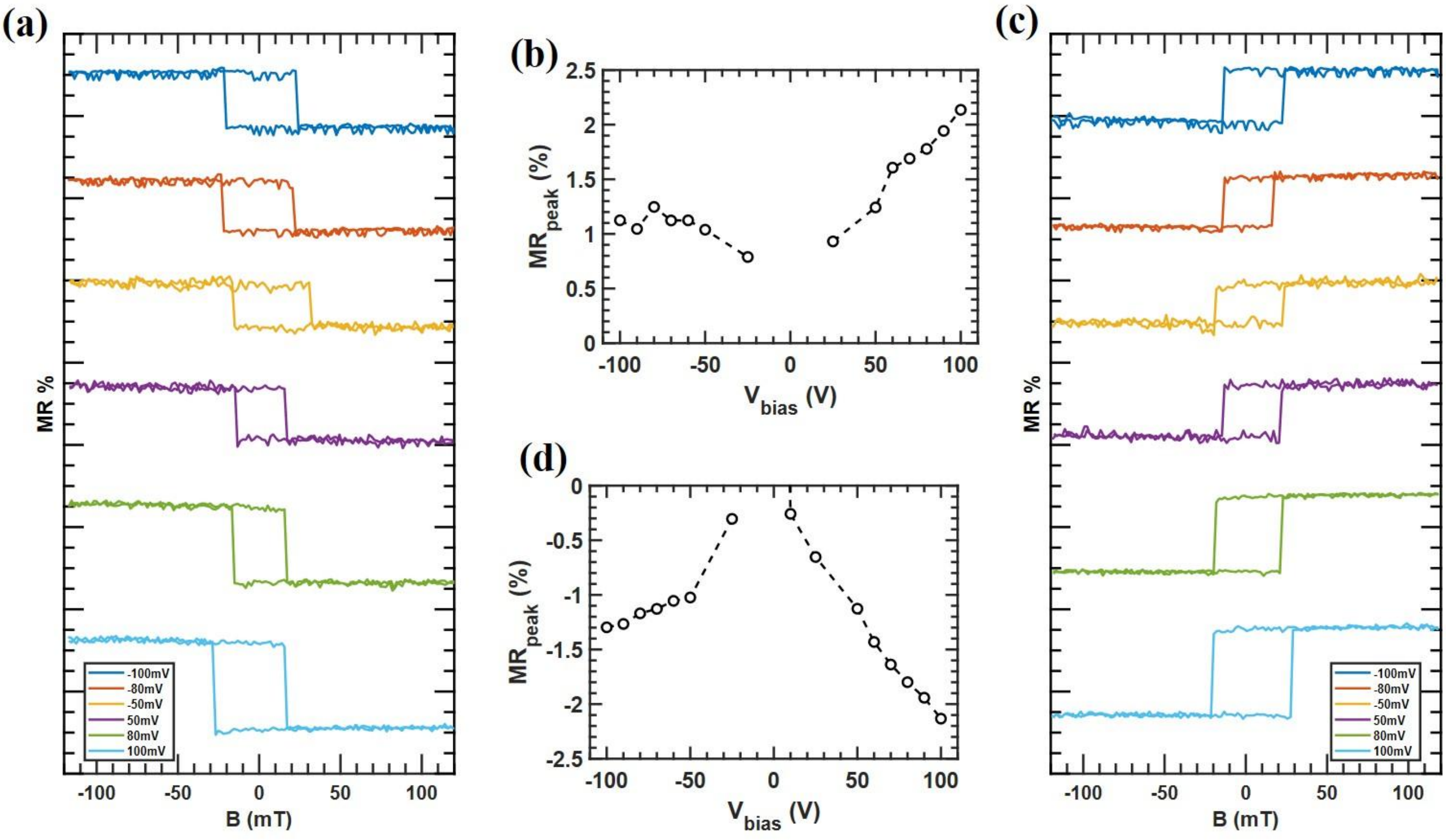

**Fig. S12: MR within the SF transition of Device C1: (a)** MR as a function of B within the SF transition region measured at different bias voltages. The Néel state is initialized by applying a −1.1 T field. The two-state MR arises from spin-dependent tunnelling between Co and CrSBr. **(b)** Extracted MR difference at B=0 T (MR peak) as a function of bias voltage. **(c)** MR as a function of B measured after −1.1 T field initialization of the Néel state, showing an opposite sign of hysteresis compared to panel (a). **(d)** Extracted MR peak values from panel (c) as a function of bias voltage.

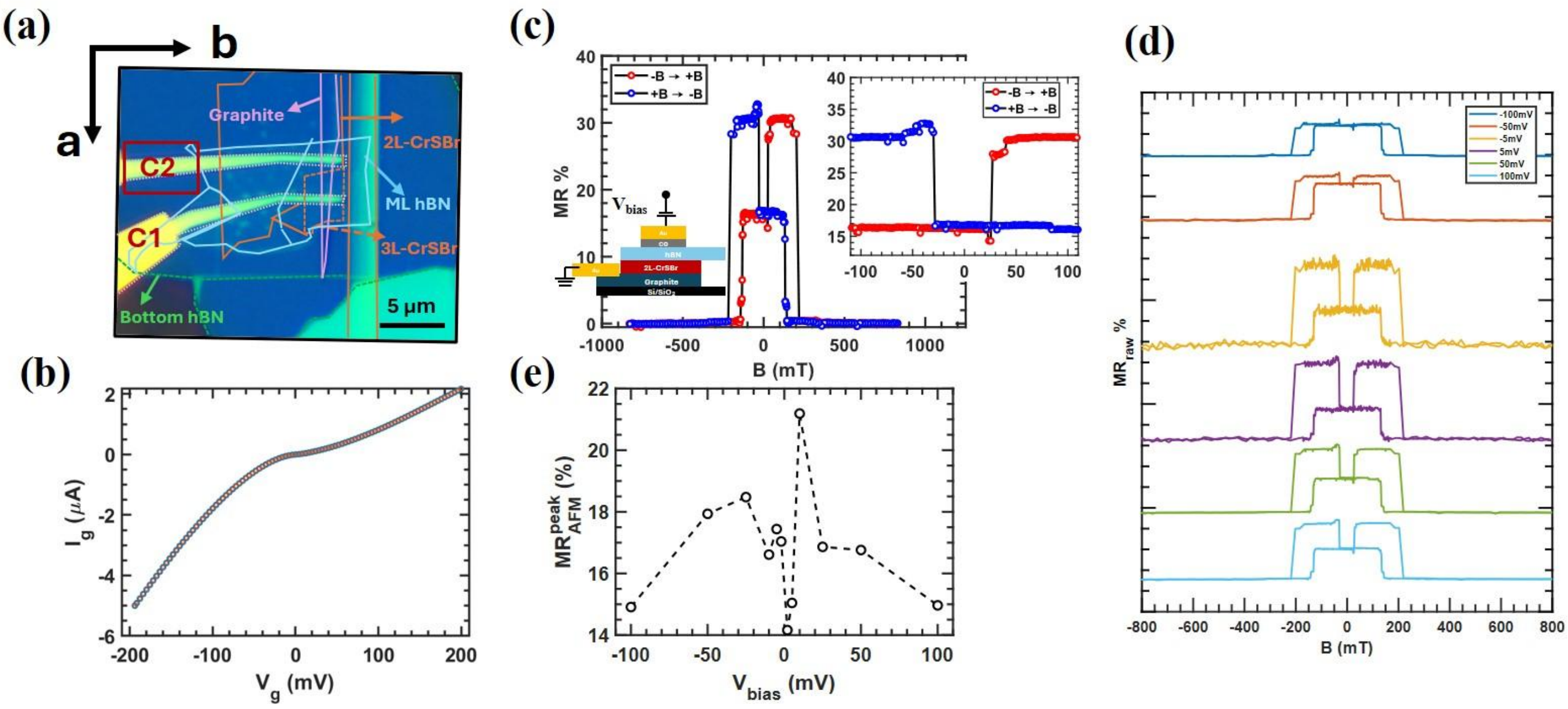

**Fig. S13: Electrical characterization of Co/hBN/2L-CrSBr Device C2: (a)** Optical image of fabricated tunnel junction device (Device C2) of Co/hBN/2L-CrSBr. Flakes are outlined and labelled. **(b)** Non-linear I-V characteristic of the device at T=10 K. **(c)** Representative plot for MR as a function of magnetic field. MR peak appearing near zero field is due to the spin dependent tunneling between Co and 2L-CrSBr. Inset showing the spin valve like behaviour as Co magnetization reversed by magnetic field. **(d)** MR vs B in complete field sweep range at various bias voltages. **(e)** MR peak values extracted from (d) as a function of bias voltage

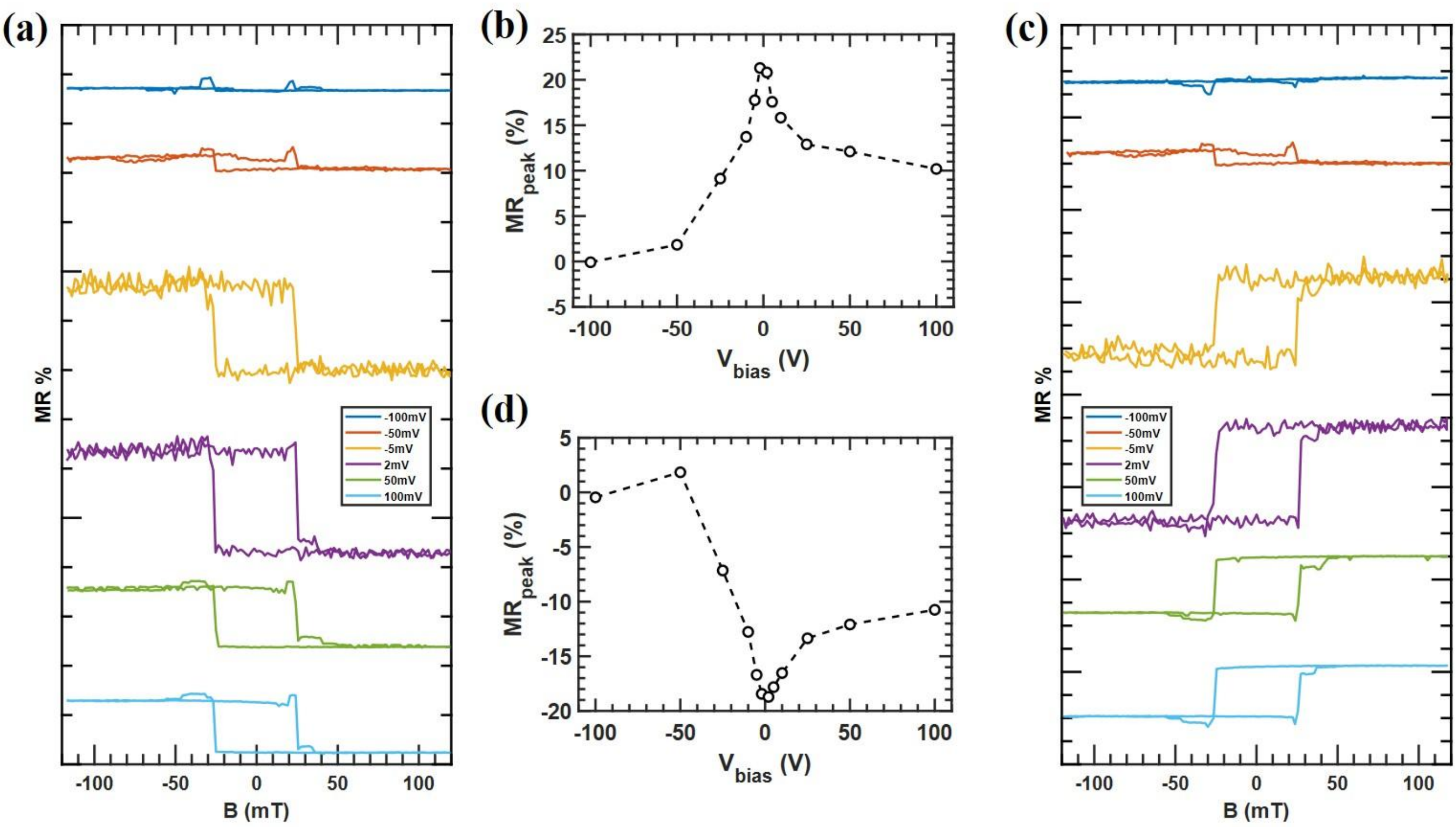

**Fig. S14: MR within the SF transition of Device C2: (a)** MR as a function of B within the SF transition region measured at different bias voltages. The Néel state is initialized by applying a −1.1 T field. The two-state MR arises from spin-dependent tunnelling between Co and CrSBr. **(b)** Extracted MR difference at B=0 T (MR peak) as a function of bias voltage. **(c)** MR as a function of B measured after −1.1 T field initialization of the Néel state, showing an opposite sign of hysteresis compared to panel (a). **(d)** Extracted MR peak values from panel (c) as a function of bias voltage.

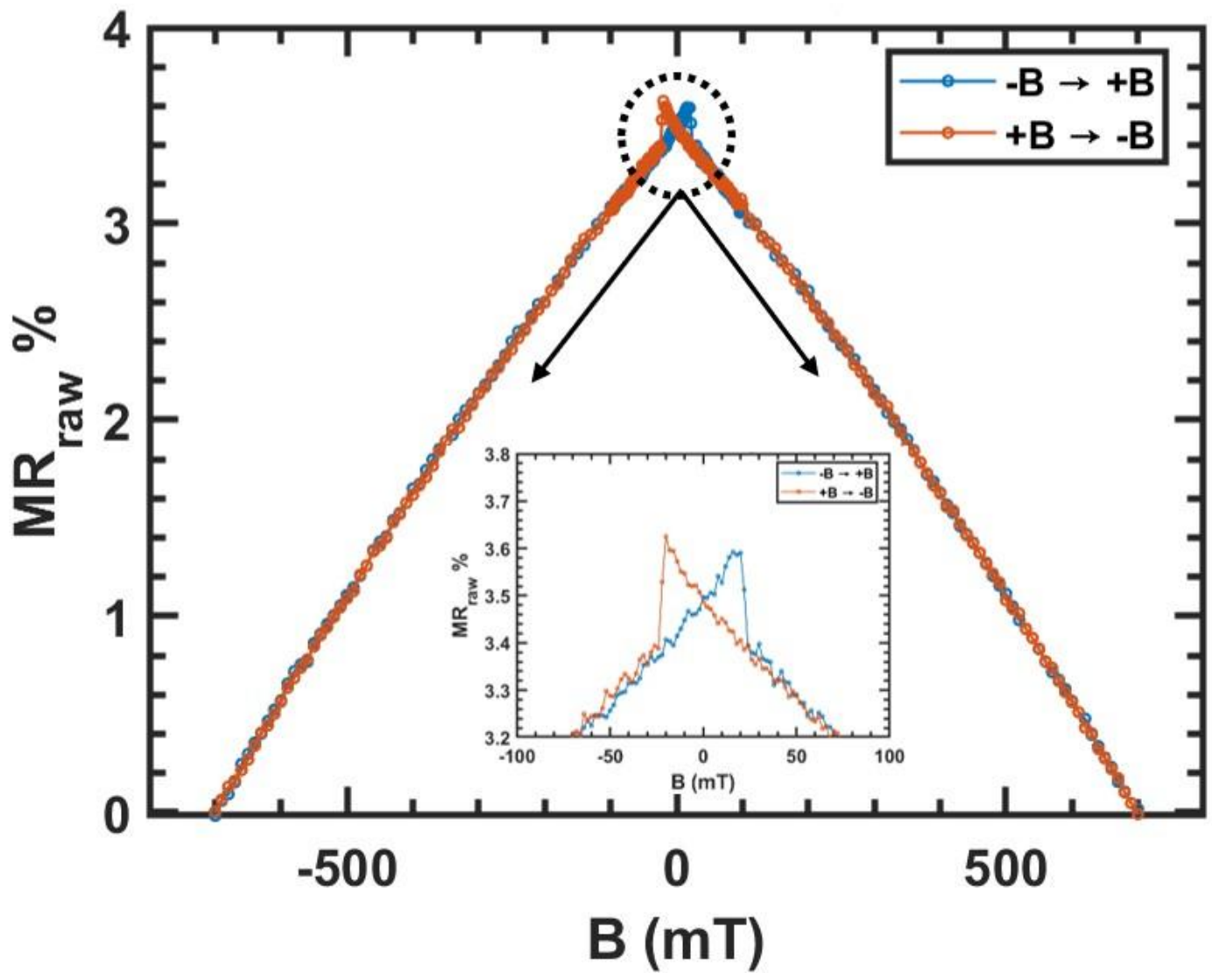

**Fig. S15. MR vs B above $T_N$ in Co/hBN/CrSBr (Device A1).** MR as a function of B measured at T=170 K, showing only the characteristic MR features of the Co electrode.

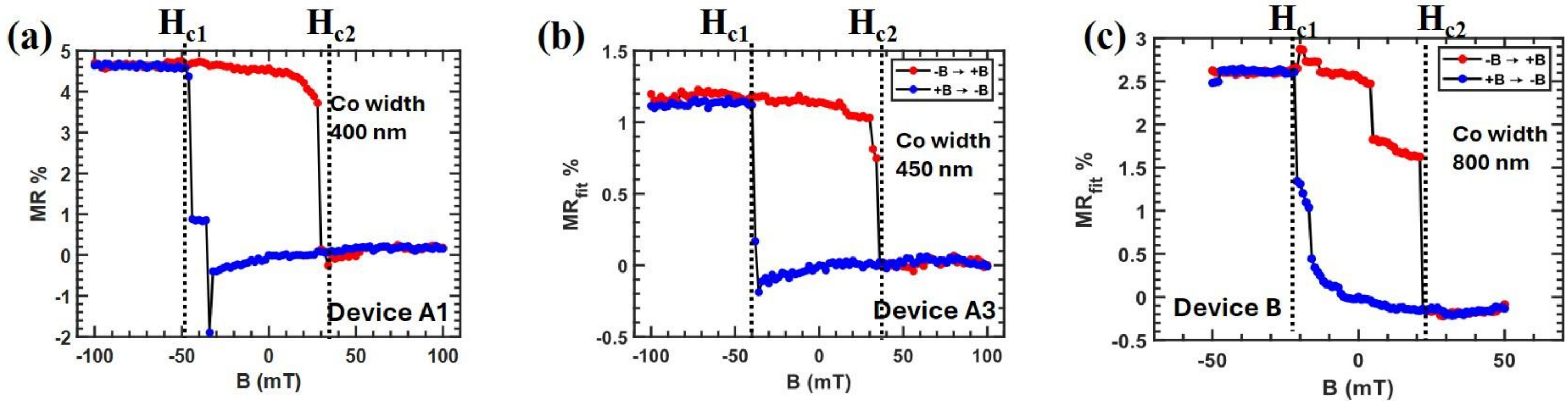

**Fig. S16: Coercivity of Co electrodes with different widths:** Magnetization reversal of the Co electrode for devices A1, A3, and B with Co widths of 400 nm, 450 nm, and 800 nm, respectively, shown in panels (a–c). The extracted coercivity values $\left(H_c = \frac{(H_{c1}+H_{c2})}{2}\right)$ are ~42 mT, ~38 mT, and ~22 mT, respectively, indicating a systematic reduction in coercivity with increasing electrode width.